\UseRawInputEncoding
\documentclass[showpacs,preprintnumbers,
amsmath,amssymb,aps,prd]{revtex4}
\usepackage{amsfonts}
\usepackage{amsfonts,graphics,epsfig,subfigure}
\usepackage{graphicx}
\usepackage{float}
\begin{document}

\title{Effects of dark energy on dynamic phase transition of charged AdS black holes}
\author{Shan-Quan Lan, Jie-Xiong Mo\footnote{mojiexiong@gmail.com(corresponding
author)}, Gu-Qiang Li, Xiao-Bao Xu}
\affiliation{Institute of Theoretical Physics, Lingnan Normal University, Zhanjiang, 524048, Guangdong, China}

\begin{abstract}
Searching for the effect of quintessence dark energy on the kinetics of black hole phase transition, we investigate in detail the dynamic phase transition of charged AdS black holes surrounded by quintessence in this paper. Based on the Gibbs free energy landscape, we obtain the analytic expression of the corresponding Gibbs free energy. As shown in $G_L-r_+$ curve at the phase transition temperature, there exist double wells with the same depth, providing further support on the finding in the former literature. By numerically solving the Fokker-Planck equation with both the initial condition and reflecting boundary condition imposed, we probe the probabilistic evolution of charged AdS black holes surrounded by quintessence. The peak denoting the initial black hole state gradually decreases while the other peak starts to grow from zero, approaching to be a stationary distribution in the long time limit with two peaks denoting the large and small black holes respectively. We also study the first passage process of charged AdS black holes surrounded by quintessence and discuss the relevant quantities. We resolve the Fokker-Planck equation by adding the absorbing boundary condition for the intermediate transition state. It is shown intuitively that the peaks located at the large (small) black hole decay very rapidly, irrespective of the initial black hole state. In all the procedures above, we have compared the cases with different choices of the state parameter of quintessence dark energy $\omega_q$. The larger $\omega_q$ is, the faster the initial black hole state decays, showing the effect of quintessence dark energy. To the best of our knowledge, it is the first probe on the influence of dark energy on the dynamic phase transition of charged AdS black hole.

\end{abstract}
   \maketitle

\section{Introduction}

    Thermodynamics especially phase transition of AdS black holes has received more and more attention in the past decade. From the perspective of the extended phase space~\cite{Kastor,Dolan1,Dolan2} including both the thermodynamic pressure and volume as variables, Kubiz\v{n}\'{a}k and Mann investigated the $P$-$V$ criticality of charged AdS black holes~\cite{Mann1} and further enhanced the analogy between charged AdS black holes and liquid-gas system observed previously by Chamblin et al.~\cite{Chamblin1,Chamblin2}. Extensive intriguing phenomena of AdS black holes have been disclosed, such as reentrant phase transition~\cite{Mann2}, triple point~\cite{Mann3,weishaowen1}, $\lambda$-line phase transition~\cite{Mann4}, $N$-fold reentrant phase transition~\cite{Hendi}, which seems to suggest that AdS black holes even share similarities with everyday thermodynamic systems. There are so many literatures that we are not able to list them all here. For reviews covering this research field, see Refs.~\cite{Mann5,Mann6}.

   Although a variety of results associated with the phase transition of AdS black holes have been reported, the underlying kinetics of black hole phase transition remained a mystery until Li and Wang et al. tried to crack down this tough problem recently~\cite{liran1,liran2}. Based on the free energy landscape, their works shed some light on both the dynamic process of the Hawing-Page phase transition~\cite{liran1} and the van der Waals like phase transition~\cite{liran2}. For Schwarzschild AdS black holes, it was found that a black hole or AdS space has the chance to transit from one phase to another due to thermal fluctuations. By studying the Fokker-Planck equation, the probabilistic evolution was probed with the mean first passage time derived~\cite{liran1}. Similarly, for Reissner-Nordstr\"{o}m-AdS (RN-AdS) black holes, the large black hole phase and the small black hole phase can transit to each other due to thermal fluctuation. And the first passage process was studied in detail~\cite{liran2}.

   These pioneer works were subsequently generalized to the case of four dimensional Gauss-Bonnet AdS black holes~\cite{liran3} and five dimensional neutral Gauss-Bonnet AdS black holes~\cite{weishaowen2}. More interestingly, Wei et al. probed the dynamic process at a black hole triple point~\cite{weishaowen3} and found that the initial large, intermediate or small black holes can switch to the other coexistent phases. Very recently, Li et al. investigated the turnover of the kinetics for the charged AdS black hole phase transition~\cite{liran4}, providing one novel way to probe the black hole microstructure.

     In the former work done by one of us~\cite{guqiang}, the $P-V$ criticality of charged AdS black holes surrounded by quintessence was investigated in the extended phase space where the cosmological constant is viewed as a variable and identified as thermodynamic pressure. It was shown that quintessence dark energy affects the critical physical quantities. Inspired by the works dealing with the kinetics of black hole phase transition mentioned above, it would certainly be interesting to probe the dynamic phase transition of charged AdS black holes surrounded by quintessence. Firstly, we can gain a dynamical picture for the phase transition of charged AdS black holes surrounded by quintessence, from which the effects of dark energy on the black hole phase transition would be more clear. Secondly, this generalization will help check whether the treatment proposed in the pioneer works of black hole dynamic phase transition is universal or not. Last but not the least, dark energy is also of enough interest in itself. Dark energy serves as a theoretical candidate to explain why the universe is expanding with acceleration. As one of the dark energy models, quintessence~\cite{Tsujikawa} may explain the late-time cosmic acceleration \cite{Fujii}-\cite{Ratra}. Both the black holes surrounded by quintessence and their thermodynamic properties have gained considerable interest~\cite{Kiselev}-\cite{linan}.

    The organization of this paper is as follows. A brief review of thermodynamics especially $P$-$V$ criticality concerning the RN-AdS black hole surrounded by quintessence will be presented in Sec.\ref{Sec2}. Dynamic phase transition of the RN-AdS black hole surrounded by quintessence will be investigated in Sec.\ref{Sec3}, where we will focus on the Gibbs free energy landscape, probabilistic evolution and the first passage process in three subsections respectively. We will end this paper with conclusions in Sec.\ref{Sec4}.

\section{A brief review on $P$-$V$ criticality of the RN-AdS black hole surrounded by quintessence}
\label{Sec2}

The metric of the RN-AdS black hole surrounded by quintessence reads~\cite{Kiselev}
\begin{equation}
ds^2=f(r)dt^2-f(r)^{-1}dr^2-r^2(d\theta^2+\sin^2\theta d\varphi^2),\label{1}\\
\end{equation}
where
\begin{equation}
f(r)=1-\frac{2M}{r}+\frac{Q^2}{r^2}-\frac{a}{r^{3\omega_q+1}}-\frac{\Lambda r^2}{3}.\label{2}\\
\end{equation}
The effect of quintessence dark energy is reflected in the fourth term of the above expression, where $\omega_q$ is the state parameter of quintessence dark energy satisfying the relation $-1<\omega_q<-1/3$. $a$ is the normalization factor which is related to the density of quintessence $\rho_q$ via
\begin{equation}
\rho_q=-\frac{a}{2}\frac{3\omega_q}{r^{3(\omega_q+1)}}.\label{3}\\
\end{equation}

The mass of the black hole $M$, the Hawking temperature $T$ and the entropy $S$ can be derived as
\begin{eqnarray}
M&=&\frac{r_+}{2}\left(1+\frac{Q^2}{r_+^2}-\frac{a}{r_+^{3\omega_q+1}}-\frac{\Lambda r_+^2}{3}\right),\label{4}\\
T&=&\frac{f'(r_+)}{4\pi}=\frac{1}{4\pi}\left(\frac{1}{r_+}-\frac{Q^2}{r_+^3}+\frac{3a\omega_q}{ r_+^{2+3\omega_q}}-r_+\Lambda\right),\label{5}\\
S&=&\int^{r_+}_0\frac{1}{T}\left(\frac{\partial M}{\partial r_+}\right)dr_+=\pi r_+^2.\label{6}
\end{eqnarray}
Treating the cosmological constant as thermodynamic pressure allows us to work in the extended phase space with the thermodynamic pressure $P$ and thermodynamic volume $V$ as
\begin{eqnarray}
P&=&-\frac{\Lambda}{8\pi},\label{7}
\\
V&=&\left(\frac{\partial M}{\partial P}\right)_{S,Q}=\frac{4\pi r_+^3}{3}.\label{8}
\end{eqnarray}

The Gibbs free energy has been obtained as~\cite{guqiang}
\begin{equation}
G=H-TS=M-TS=\frac{3Q^2}{4r_+}+\frac{r_+}{4}-\frac{2P\pi r_+^3}{3}-\frac{a(2+3\omega_q)}{4r_+^{3\omega_q}}.\label{9}
\end{equation}
Note that the mass has been interpreted as enthalpy in the extended phase space~\cite{Kastor}.

$P$-$V$ criticality was discussed in detail in Ref.~\cite{guqiang}. It is worth mentioning that we can obtain the analytic expression of critical quantities as follow when $\omega_q=-2/3$
\begin{equation}
v_c=2\sqrt{6}Q,\;\;T_c=\frac{\sqrt{6}}{18\pi Q}-\frac{a}{2\pi},\;\;P_c=\frac{1}{96\pi Q^2},\label{10}
\end{equation}
where $v_c, T_c, P_c$ denote the critical specific volume, critical Hawking temperature and critical thermodynamic pressure respectively. When $\omega_q\neq-2/3$, critical quantities can be obtained numerically for different choices of parameters.

\section{Dynamic phase transition of the RN-AdS black hole surrounded by quintessence}
\label{Sec3}

\subsection{Gibbs free energy landscape for the RN-AdS black hole surrounded by quintessence}

In the $P$-$V$ criticality research~\cite{Mann1}, Gibbs free energy plays a very important role by identifying the first order phase transition via the classical swallowtail behavior. In the pioneer works of black hole dynamic phase transition~\cite{liran1,liran2}, Gibbs free energy also plays a crucial role. The analysis of dynamic phase transition is based on the so-called "Gibbs free energy landscape", where Gibbs free energy $G_L$ has a similar definition as Eq. (\ref{9}) other than the replacement of the Hawking temperature $T$ with the temperature of the ensemble $T_E$. It was argued that $G_L$ describes a real black hole only when $T_E=T$~\cite{liran1,liran2}. It was further proved that only the extremal points of $G_L$ make sense with local minimum and maximum corresponding to the stable and unstable phases respectively. And the small-large black hole phase transition takes place when the double wells of $G_L$ have the same depth~\cite{weishaowen2}.

Utilizing Eqs. (\ref{4}), (\ref{6}) and (\ref{7}), we can obtain the expression of Gibbs free energy $G_L$ on the Gibbs free energy landscape as
\begin{equation}
G_L=H-T_ES=\frac{r_+}{2}\left(1+\frac{Q^2}{r_+^2}-\frac{a}{r_+^{3\omega_q+1}}+\frac{8\pi P r_+^2}{3}-2\pi r_+T_E\right).\label{11}
\end{equation}

\begin{figure}[H]
\centerline{\subfigure[]{\label{w23a0.1T0.02gt}
\includegraphics[width=8cm,height=6cm]{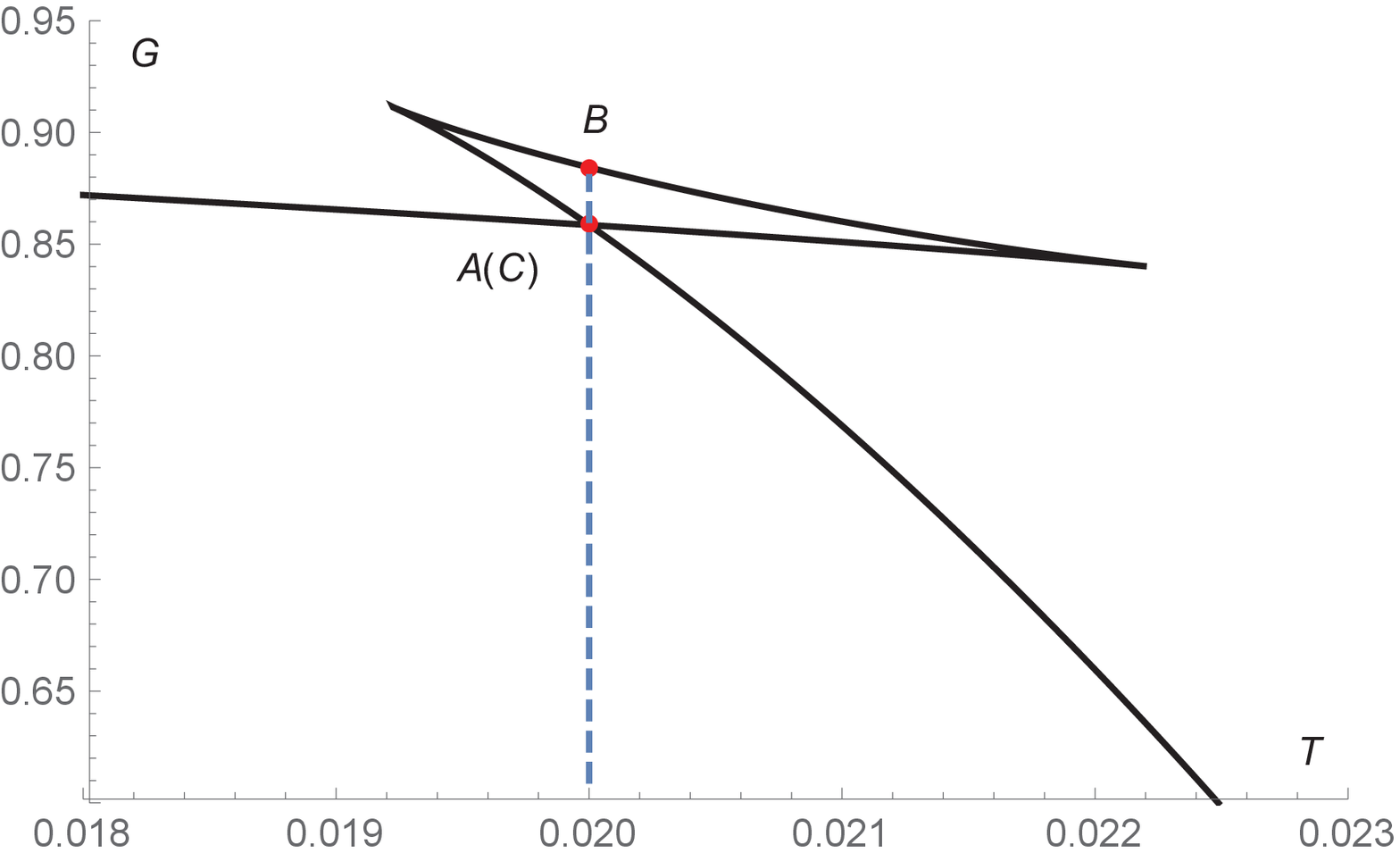}}
\subfigure[]{\label{w23a0.1T0.02gr}
\includegraphics[width=8cm,height=6cm]{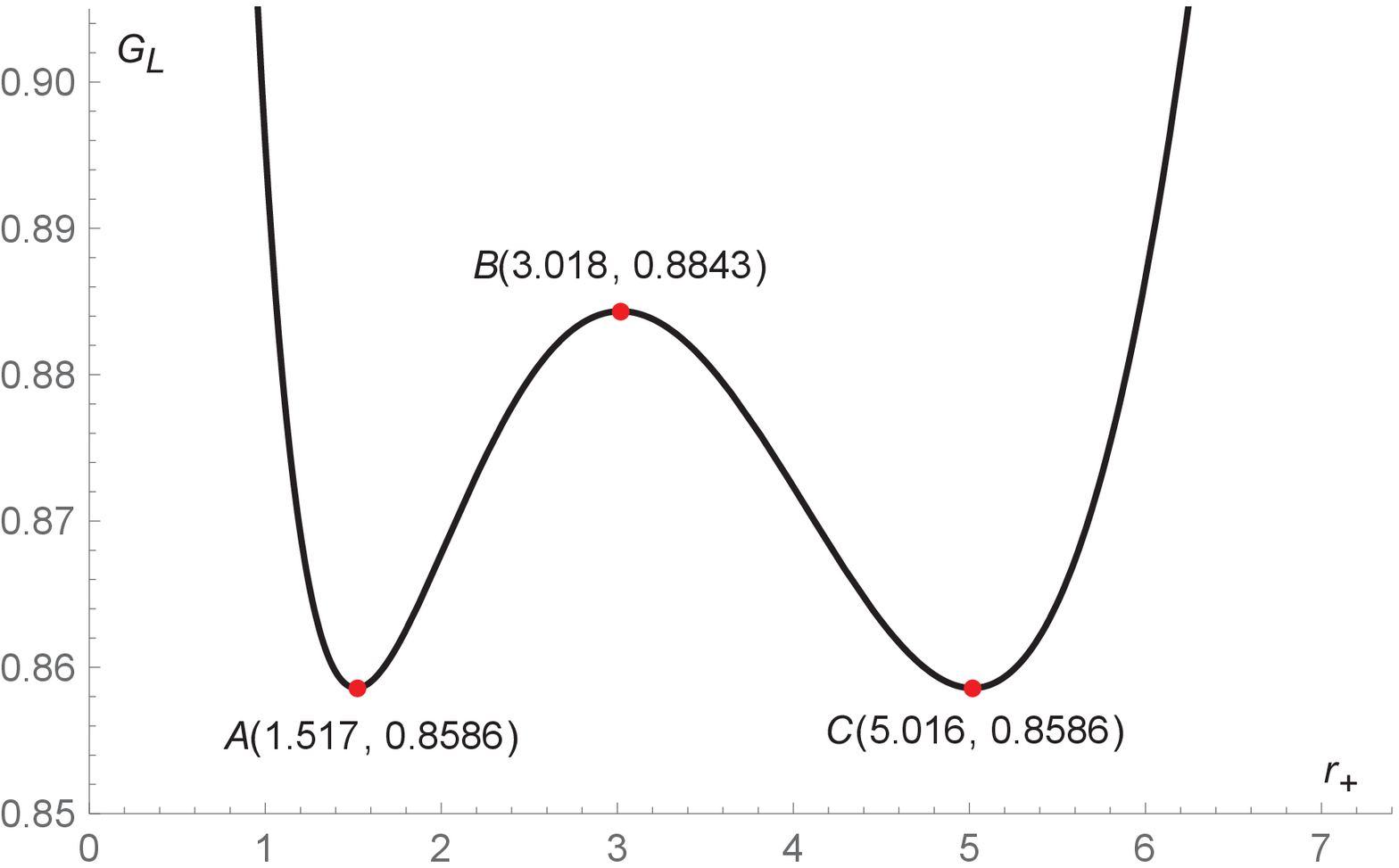}}}
 \caption{Gibbs free energy for the RN-AdS black hole surrounded by quintessence with $\omega_q=-2/3, a=0.1, P=0.0206$. (a) $G-T$ graph (b) $G_L$ vs $r_+$ when $T_E=0.02$}
\label{fg0}
\end{figure}

To gain an intuitive understanding, we present a specific example in Fig.\ref{fg0}. From the $G-T$ graph on the left side, it can be witnessed that there exist three branches. Namely, the large black hole branch, the intermediate black hole branch and the small black hole branch. The swallow tail behavior which is characteristic of first order phase transition appears when the pressure is fixed at a value lower than the critical thermodynamic pressure. We can read off the phase transition temperature from the intersection point between the large black hole branch and the small black hole branch as $0.02$. Note that in the following investigation of
dynamic phase transition, we will consider the case $T_E=0.02$ for consistency. From the right graph, one can see clearly there exist double wells in the $G_L-r_+$ curve. And these two wells have the same depth, providing further support on the former findings concerning the Gauss-Bonnet black holes~\cite{weishaowen2}.

\subsection{Probabilistic evolution of the RN-AdS black hole surrounded by quintessence}

As a useful tool to probe the probabilistic evolution of the black hole, Fokker-Planck equation reads~\cite{Zwanzig,Lee1,Lee2,wang1,Bryngelson}
\begin{equation}
\frac{\partial \rho(r,t)}{\partial t}=D\frac{\partial}{\partial r}\left(e^{-\beta G_L(T,P,r)}\frac{\partial}{\partial r}\big(e^{\beta G_L(T,P,r)}\rho(r,t)\big)\right),\label{12}
\end{equation}
where $\rho(r,t)$ presents the probability distribution picture of black hole phases after the thermal fluctuation based on the Gibbs free energy landscape characterized by $G_L$. In the Fokker-Planck equation, $D$ is the diffusion coefficient with its definition as $D=\frac{k_B T}{\zeta}$, where $\zeta$ and $k_B$ is the dissipation coefficient and Boltzman constant respectively. And the parameter $\beta=\frac{1}{k_B T}$. Both $k_B$ and $\zeta$ can be set to one without loss of generality. Note that we have denoted the black hole horizon radius $r_+$ as $r$ for simplicity and we will use this notation in the following.

\begin{figure}[H]
\centerline{\subfigure[]{\label{w23a0.1T0.02L1}
\includegraphics[width=8cm,height=6cm]{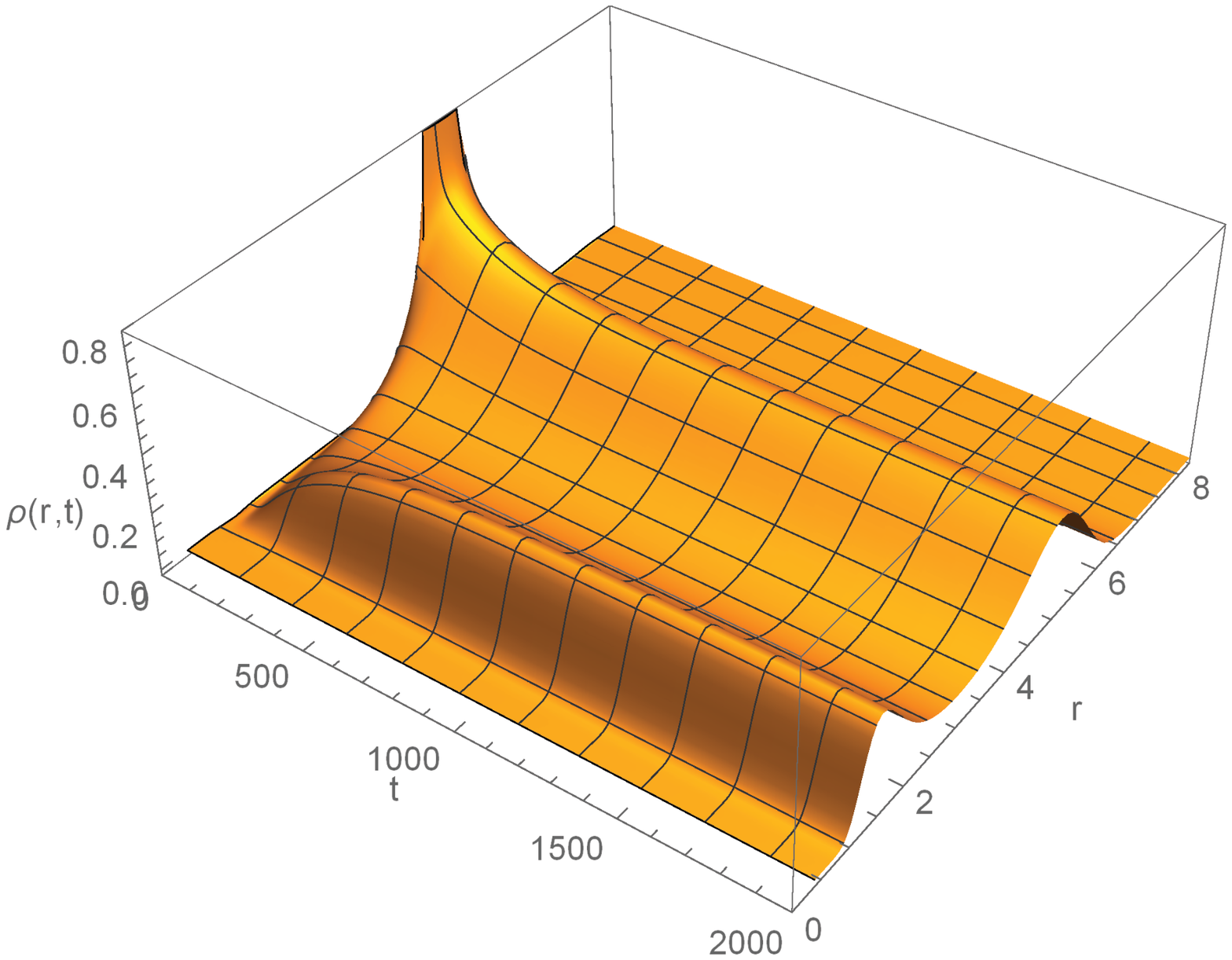}}
\subfigure[]{\label{w23a0.1T0.02S1}
\includegraphics[width=8cm,height=6cm]{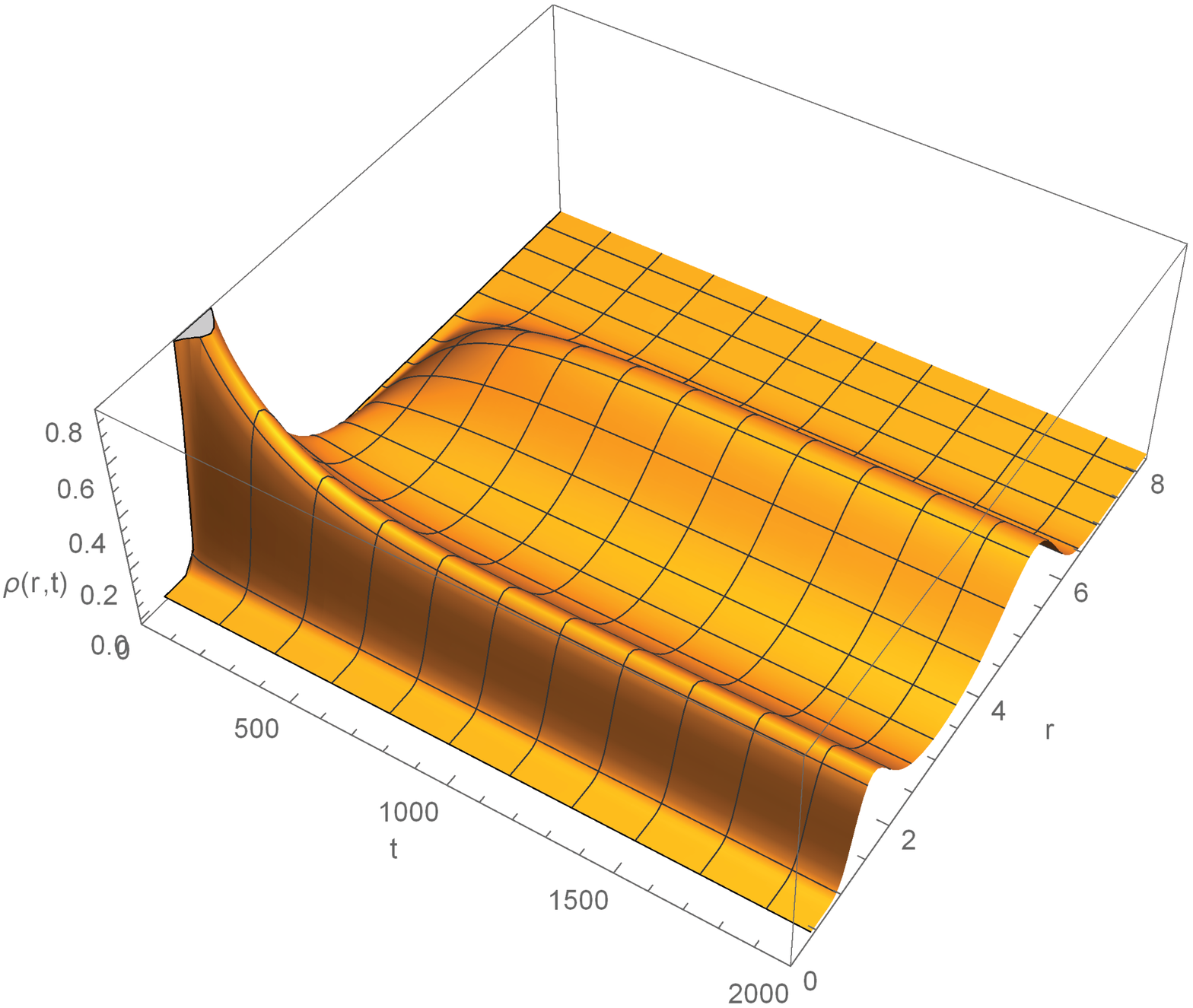}}}
\centerline{\subfigure[]{\label{w0.9a0.1T0.02L1}
\includegraphics[width=8cm,height=6cm]{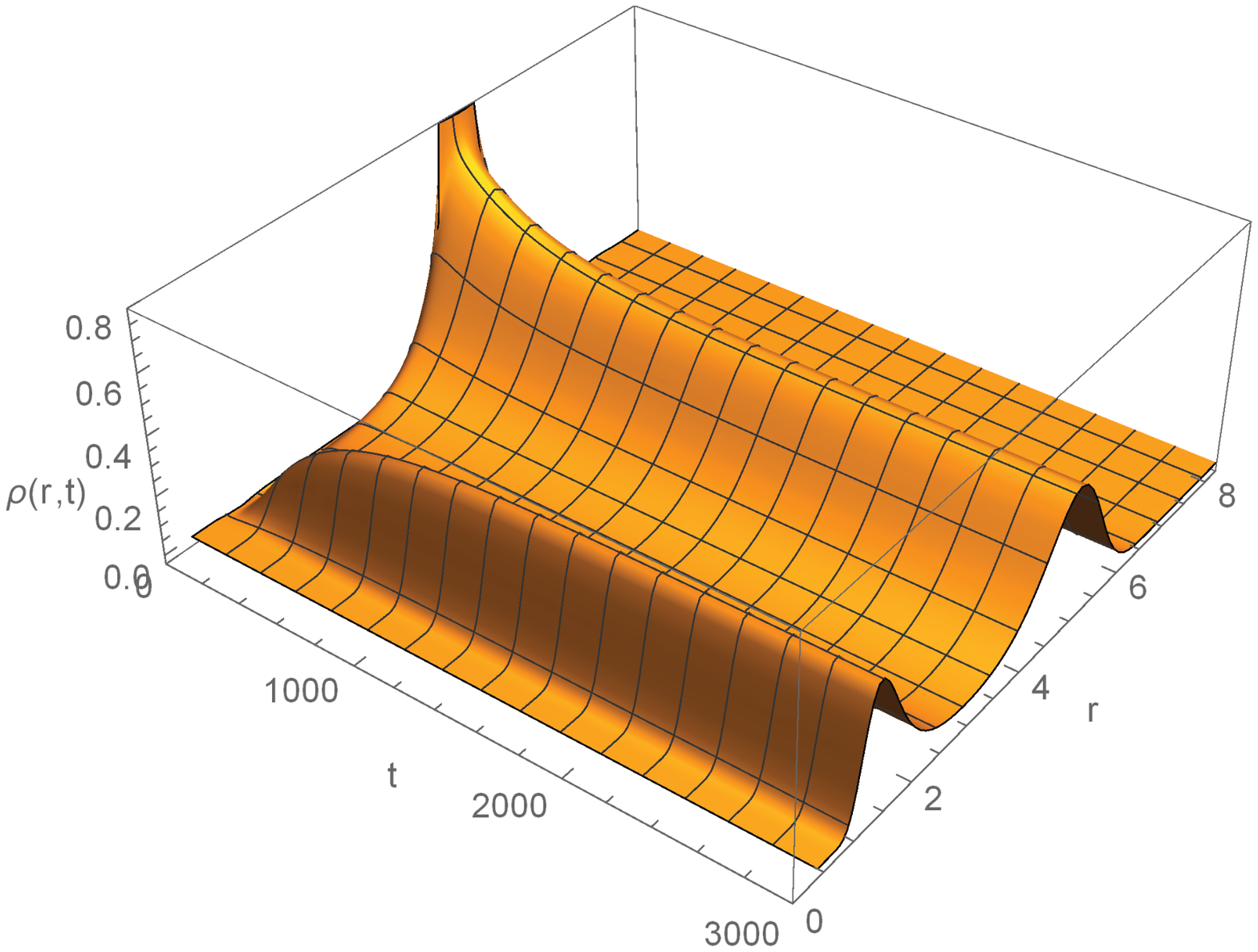}}
\subfigure[]{\label{w0.9a0.1T0.02S1}
\includegraphics[width=8cm,height=6cm]{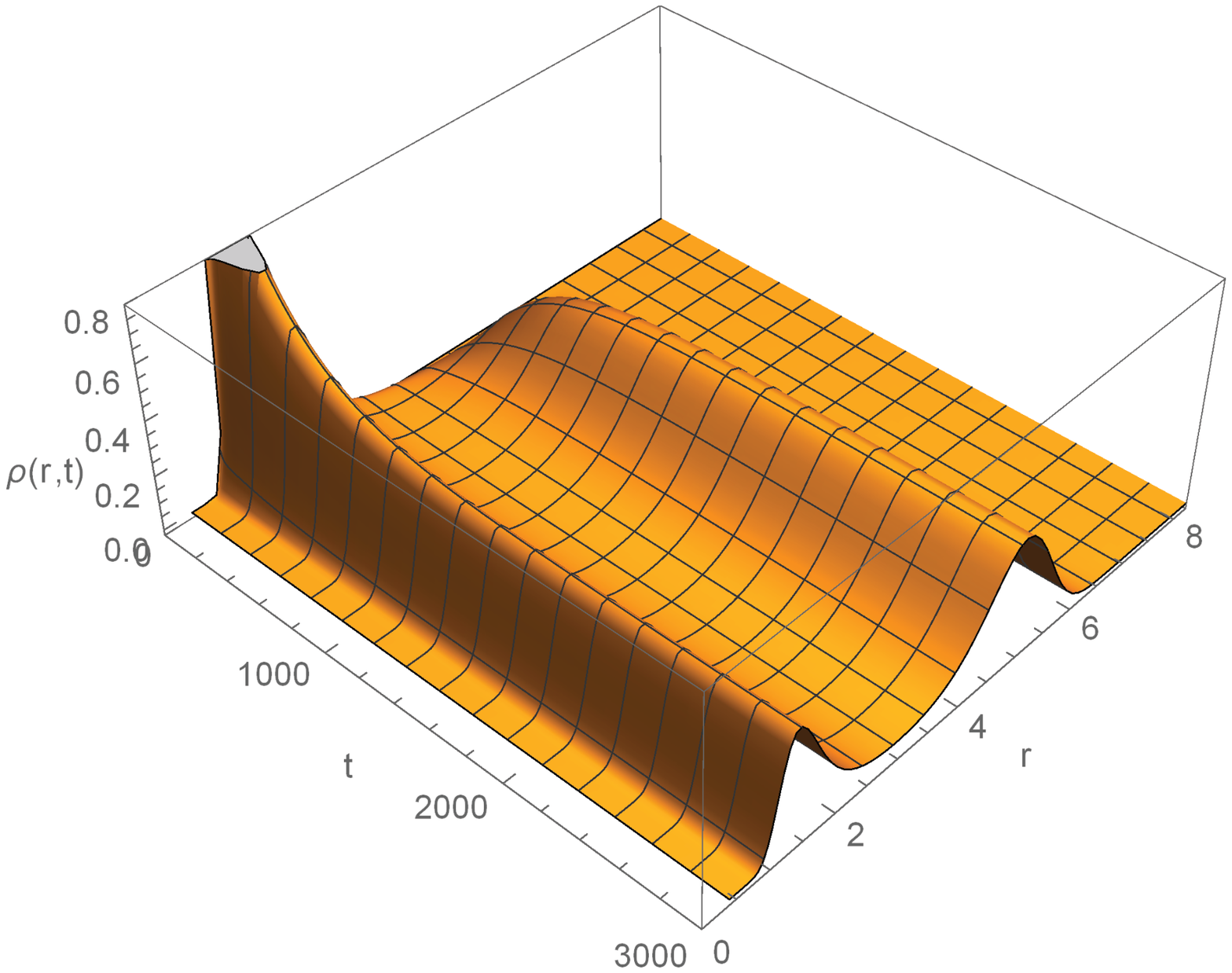}}}
 \caption{Time evolution of $\rho(r,t)$ for the RN-AdS black hole surrounded by quintessence with $a=0.1, T_E=0.02$. (a) $\omega_q=-2/3$ (b) $\omega_q=-2/3$ (c) $\omega_q=-0.9$ (d) $\omega_q=-0.9$. For the left two graphs, the initial condition is chosen as Gaussian wave pocket located at the large black hole state while for the right two graphs, the initial condition is chosen as Gaussian wave pocket located at the small black hole state.}
\label{fg1}
\end{figure}

We can obtain the time evolution picture of the probability distribution $\rho(r,t)$ by numerically solving Eq. (\ref{12}) with both the initial condition and the reflecting boundary condition ($\beta G'(r)\rho(r,t)+\rho'(r,t)\big|_{r=r_0}=0$~\cite{liran2}) imposed. Note that we consider the reflecting boundary condition at $r=0.3$ and $r=8.3$ to avoid the numerical instability since the Gibbs free energy is divergent at $r=0$ and $r=+\infty$. As shown in Fig.\ref{fg1}, we plot the evolution of $\rho(r,t)$ for the RN-AdS black hole surrounded by quintessence with $a=0.1, T_E=0.02$. In the top and bottom rows, the state parameter of quintessence dark energy $\omega_q$ is chosen as $-2/3$ and $-0.9$ respectively for comparison to probe the effect of dark energy on the evolution. For the left two graphs, the initial condition is chosen as Gaussian wave pocket located at the large black hole state ($\rho(r,0)=\frac{1}{\sqrt{\pi}b}e^{-(r-r_l)^2/b^2}$~\cite{liran2}, where $r_l$ denotes the horizon radius of large black hole) while for the right two graphs, the initial condition is chosen as Gaussian wave pocket located at the small black hole state ($\rho(r,0)=\frac{1}{\sqrt{\pi}b}e^{-(r-r_s)^2/b^2}$~\cite{liran2}, where $r_s$ denotes the horizon radius of small black hole). Here, we use the notation $b$ instead of $a$ in Ref.~\cite{liran2} to avoid the confusion with the normalization factor $a$ which is related to the density of quintessence.

\begin{figure}[H]
\centerline{\subfigure[]{\label{w23a0.1T0.02L2}
\includegraphics[width=8cm,height=6cm]{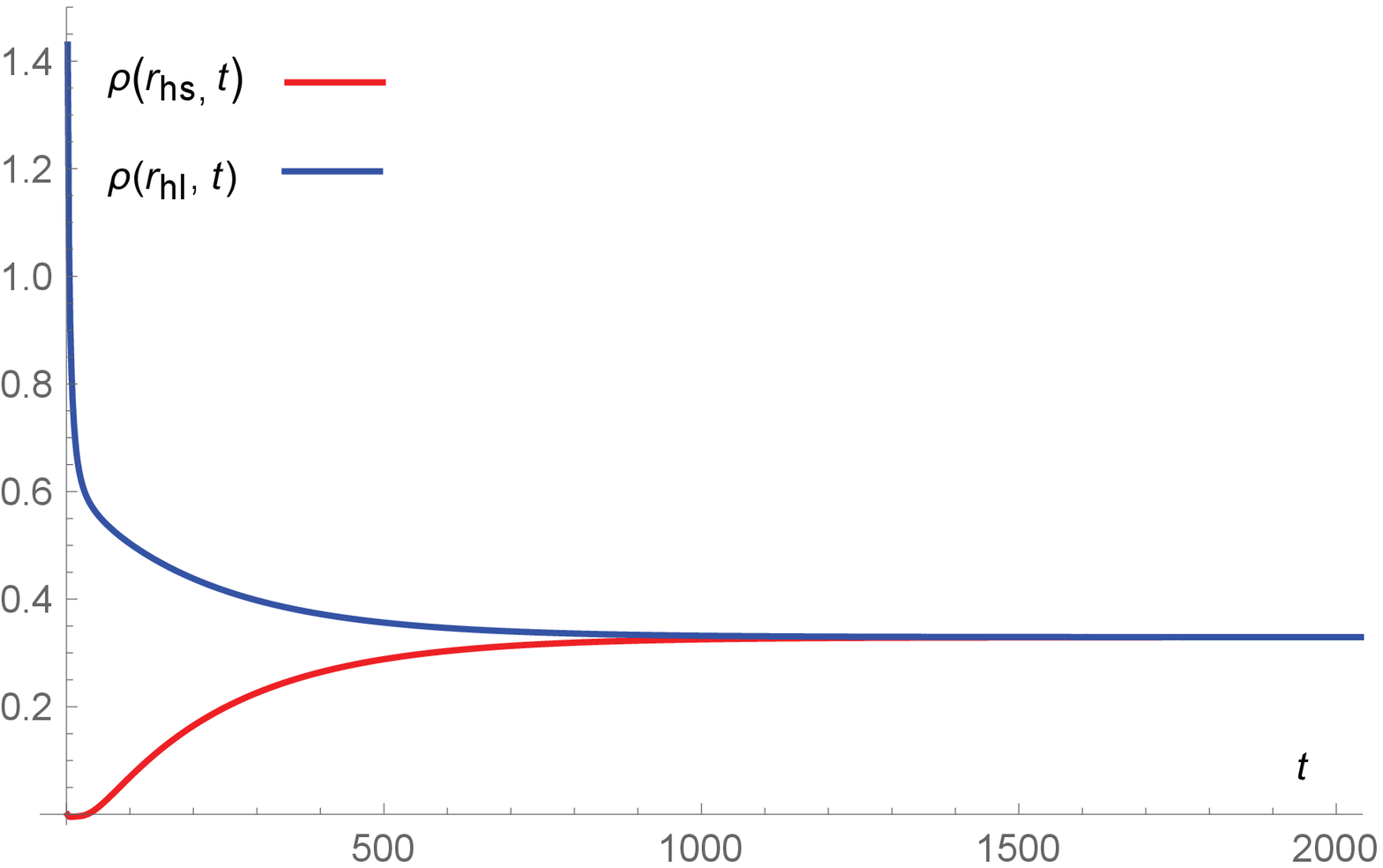}}
\subfigure[]{\label{w23a0.1T0.02S2}
\includegraphics[width=8cm,height=6cm]{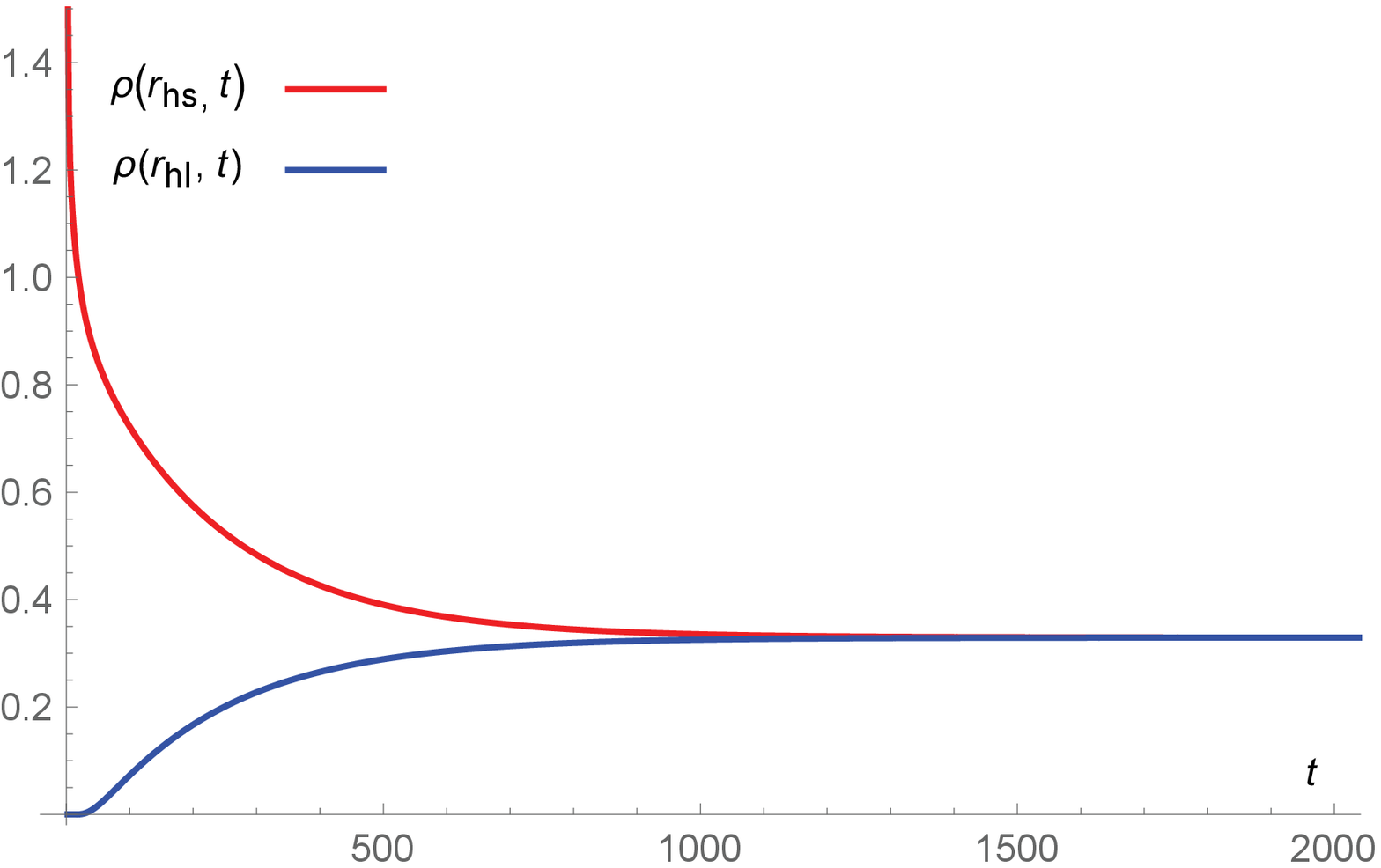}}}
\centerline{\subfigure[]{\label{w0.9a0.1T0.02L2}
\includegraphics[width=8cm,height=6cm]{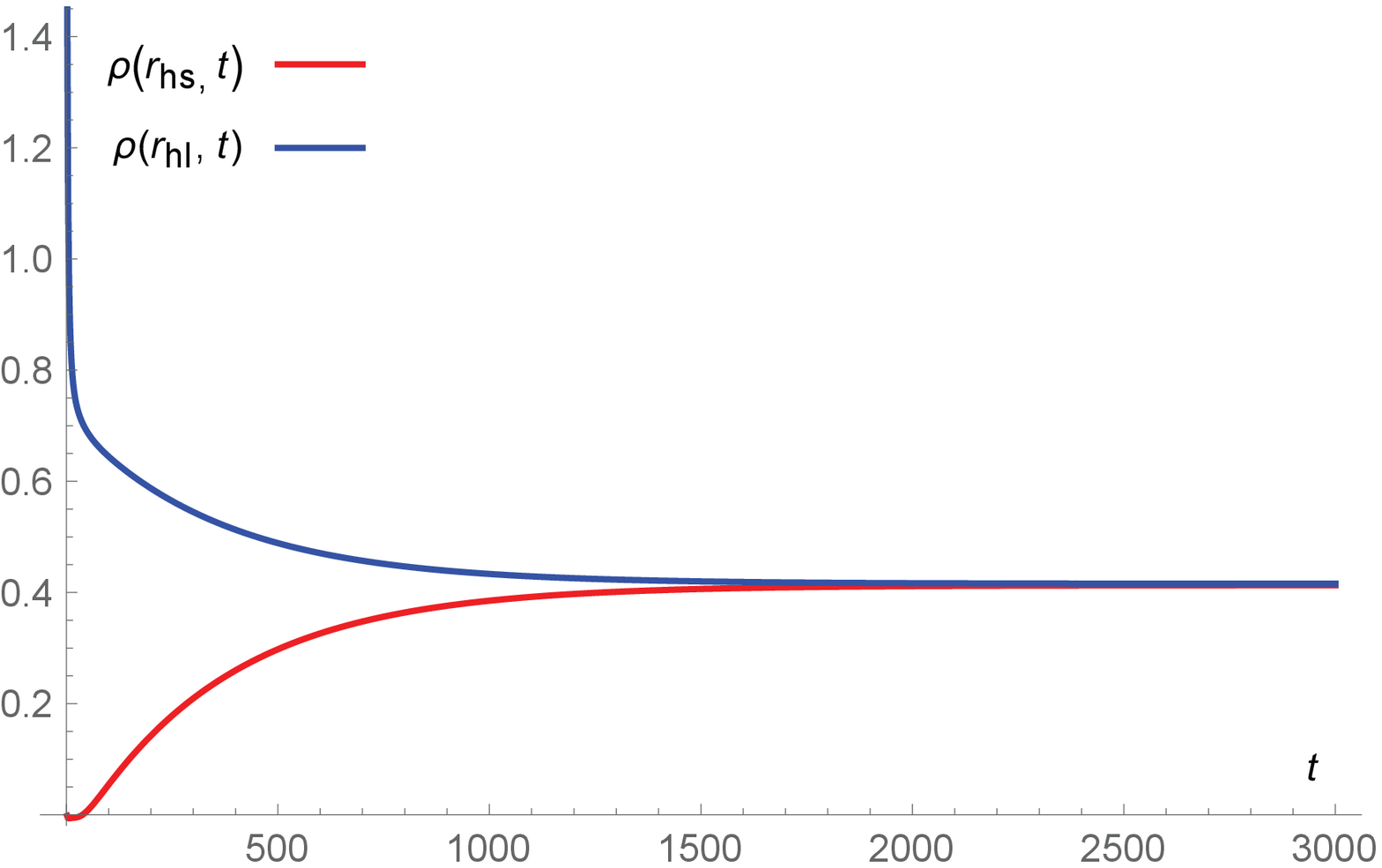}}
\subfigure[]{\label{w0.9a0.1T0.02S2}
\includegraphics[width=8cm,height=6cm]{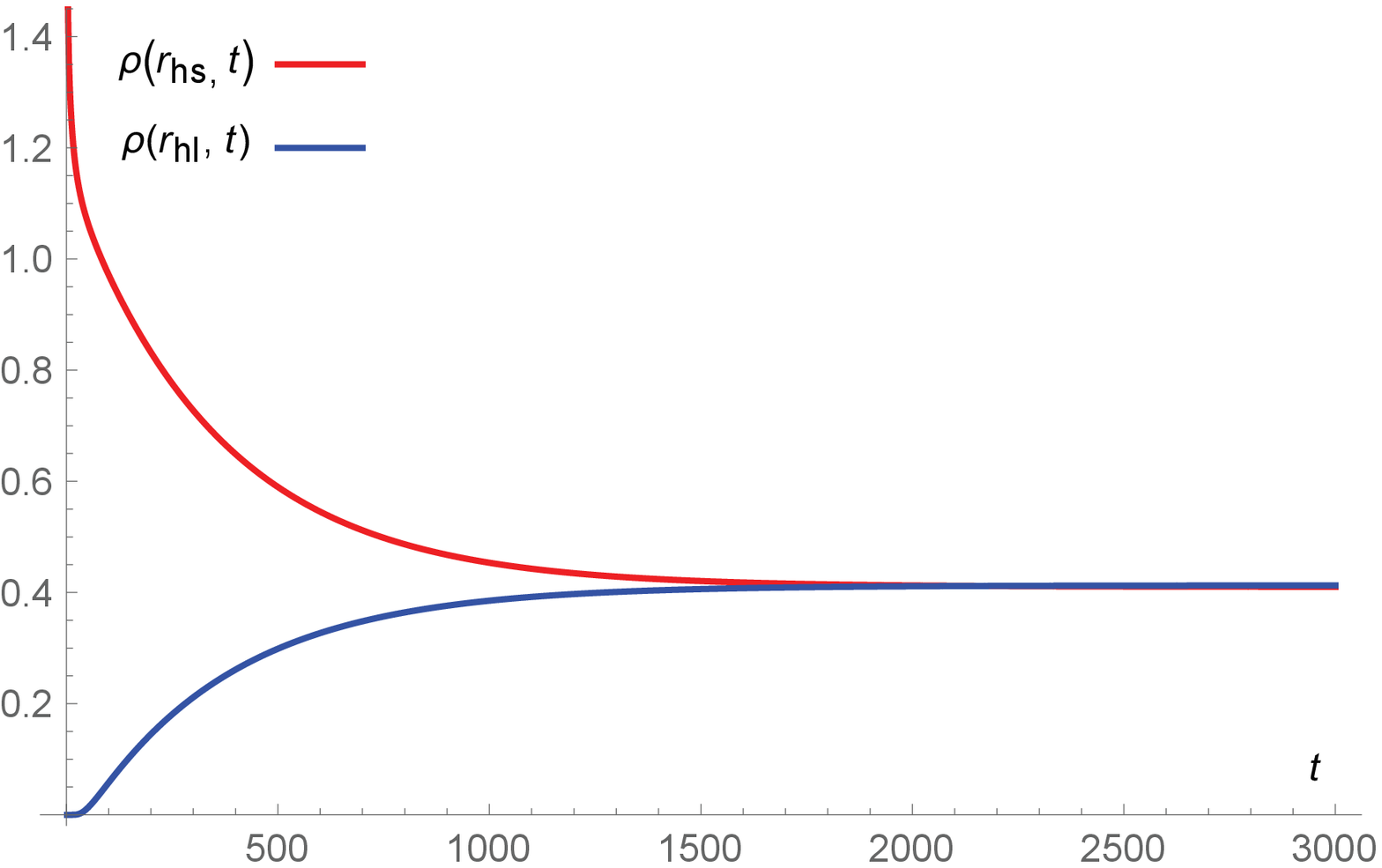}}}
 \caption{Time evolution of $\rho(r_l,t)$ and $\rho(r_s,t)$ for the RN-AdS black hole surrounded by quintessence with $a=0.1, T_E=0.02$. (a) $\omega_q=-2/3$ (b) $\omega_q=-2/3$ (c) $\omega_q=-0.9$ (d) $\omega_q=-0.9$. For the left two graphs, the initial condition is chosen as Gaussian wave pocket located at the large black hole state while for the right two graphs, the initial condition is chosen as Gaussian wave pocket located at the small black hole state.}
\label{fg2}
\end{figure}

Fig.\ref{fg1} allows one to gain an overall picture of how the probability distribution $\rho(r,t)$ evolves. For a precise understanding, we plot the time evolution of both $\rho(r_l,t)$ and $\rho(r_s,t)$ in Fig.\ref{fg2}. The parameters here are chosen the same as Fig.\ref{fg1}. In the top and bottom rows, $\omega_q$ is chosen as $-2/3$ and $-0.9$ respectively. Also, the initial condition is chosen as Gaussian wave pocket located at the large black hole state for the left two graphs while the initial condition is chosen as Gaussian wave pocket located at the small black hole state for the right two graphs.

As can be seen from Figs.\ref{w23a0.1T0.02L1} and \ref{w0.9a0.1T0.02L1} where the initial state is the large black hole, the peak denoting the large black hole gradually decreases while the other peak denoting the small black hole starts to grow from zero. And $\rho(r,t)$ approaches to be a stationary distribution in the long time limit with two peaks denoting the large and small black holes respectively. This clearly exhibits the dynamic process of how the initial large black hole evolves into the small black hole. More specifically, the evolution of both the $\rho(r_l,t)$ and $\rho(r_s,t)$ can be reflected in Figs.\ref{w23a0.1T0.02L2} and \ref{w0.9a0.1T0.02L2}. At first, $\rho(r_l,t)$ is finite while $\rho(r_s,t)$ equals to zero. During the evolution, $\rho(r_l,t)$ decreases while $\rho(r_s,t)$ increases. Finally, they approach to the same value ($0.3293$ for the case of $\omega_q=-2/3$ and $0.4167$ for the case of $\omega_q=-0.9$), consistent with the conclusion in Ref~\cite{weishaowen2} that the final probability of the large black hole equals to that of the small black hole because their Gibbs free energy equal to each other. Similar analysis is applicable when the initial black hole state is small black hole.

Comparing the top two graphs with the bottom two graphs in both Fig.\ref{fg1} and Fig.\ref{fg2}, we can see that the initial black hole state evolves more quickly in the case of $\omega_q=-2/3$ than in the case $\omega_q=-0.9$. Considering $\omega_q$ is the state parameter of quintessence dark energy, the phenomenon that different choices of $\omega_q$ lead to different evolution rate reflects the effect of quintessence dark energy on the dynamic phase transition of black holes.

\subsection{First passage process for the phase transition of the RN-AdS black hole surrounded by quintessence}

First passage process for black hole phase transition describes such a process that the initial small (large) black hole state reaches the intermediate transition state for the first time. This process can be interpreted intuitively from the Gibbs free energy landscape that for the first time the state at the well of $G_L$ reaches the state at the barriers of $G_L$. The corresponding time for this process is called the first passage time.

\begin{figure}[H]
\centerline{\subfigure[]{\label{w23a0.1T0.02L3}
\includegraphics[width=8cm,height=6cm]{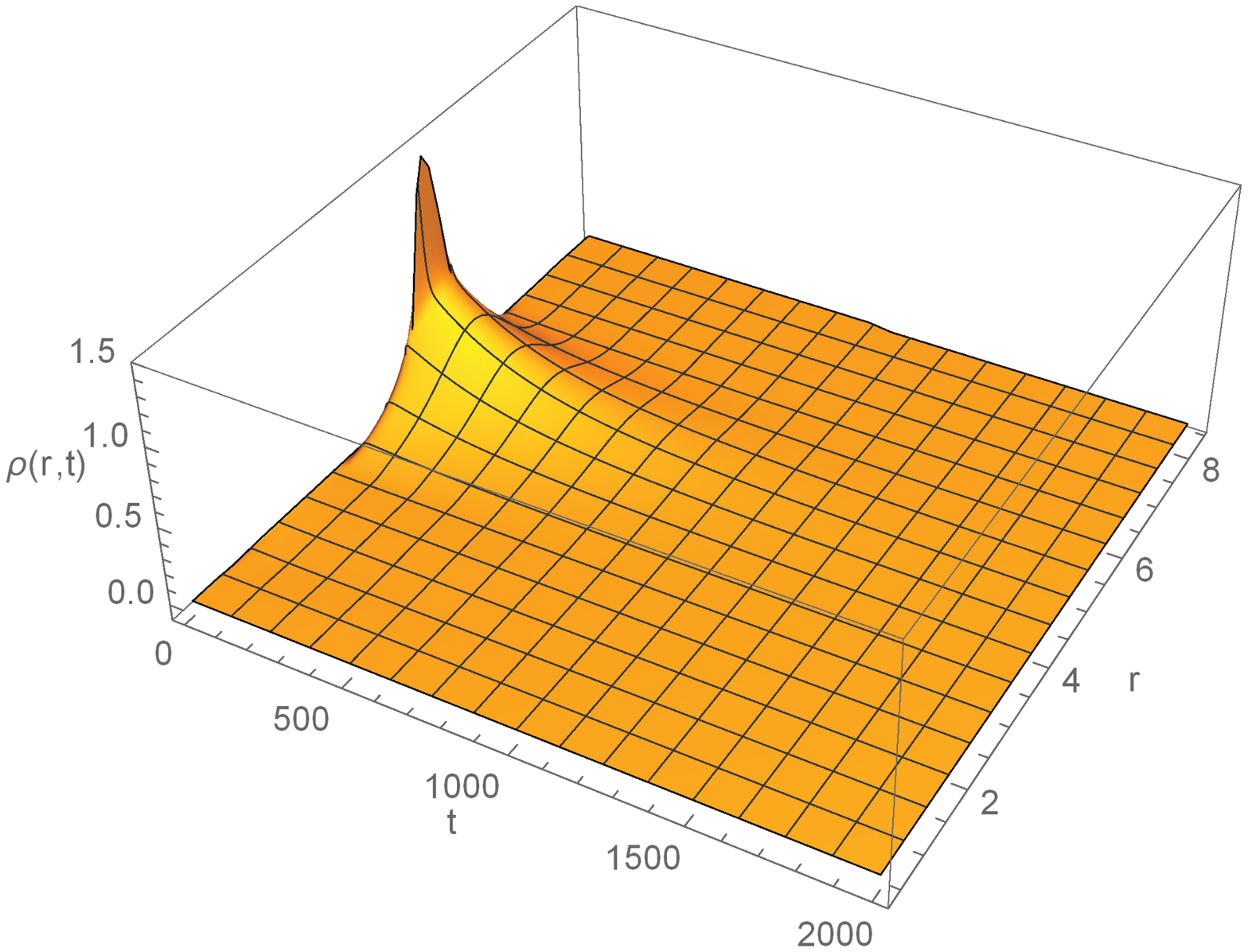}}
\subfigure[]{\label{w23a0.1T0.02S3}
\includegraphics[width=8cm,height=6cm]{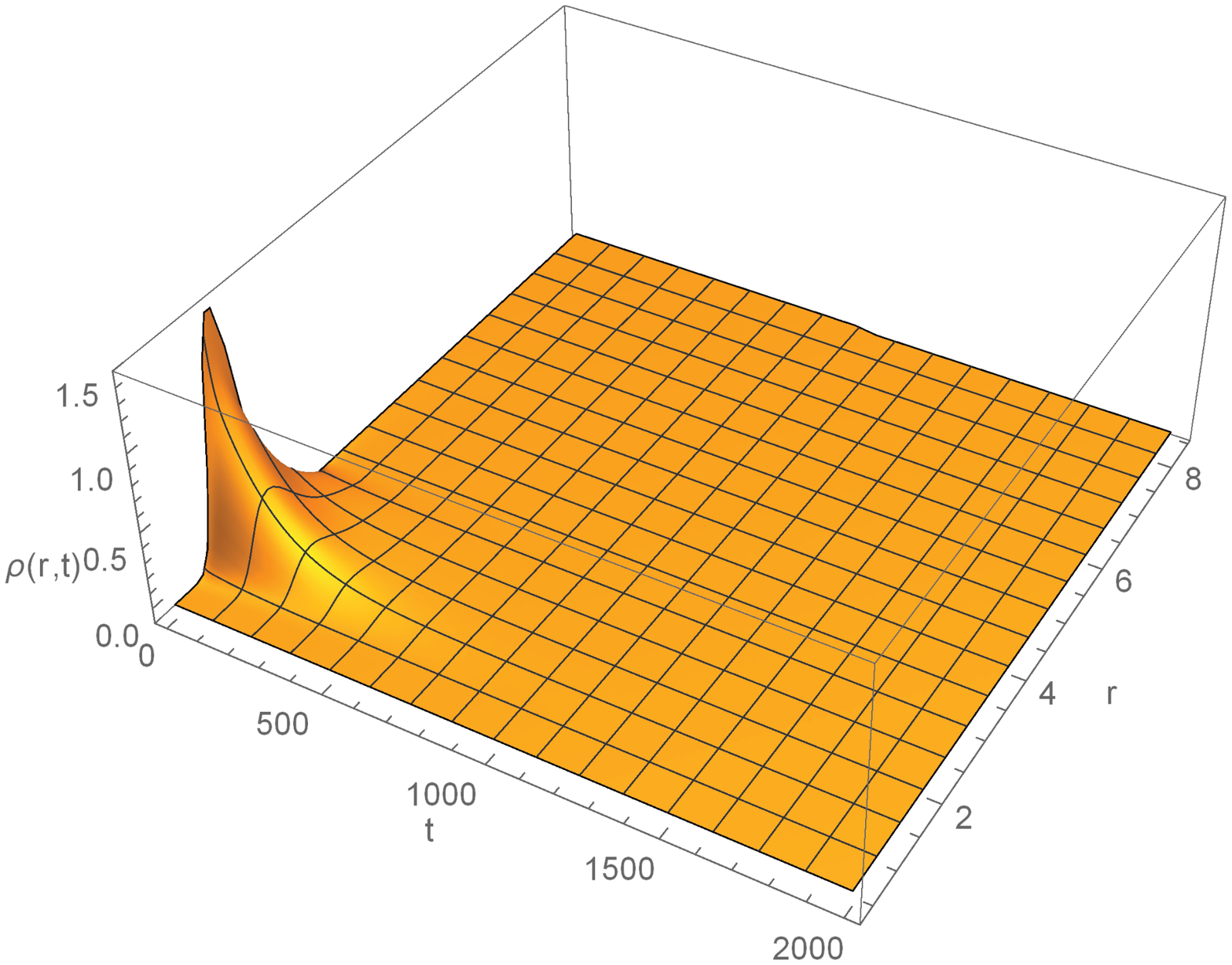}}}
\centerline{\subfigure[]{\label{w0.9a0.1T0.02L3}
\includegraphics[width=8cm,height=6cm]{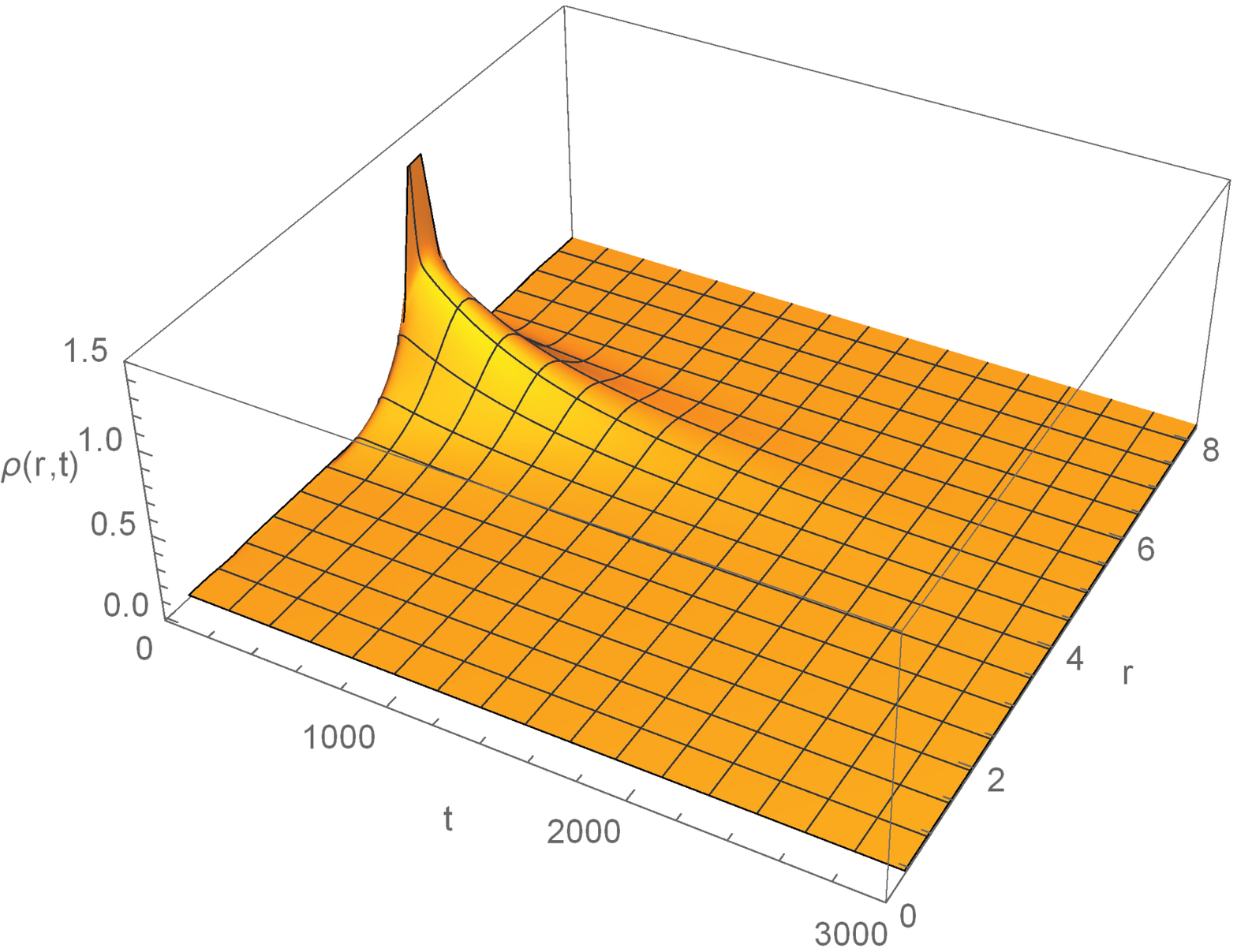}}
\subfigure[]{\label{w0.9a0.1T0.02S3}
\includegraphics[width=8cm,height=6cm]{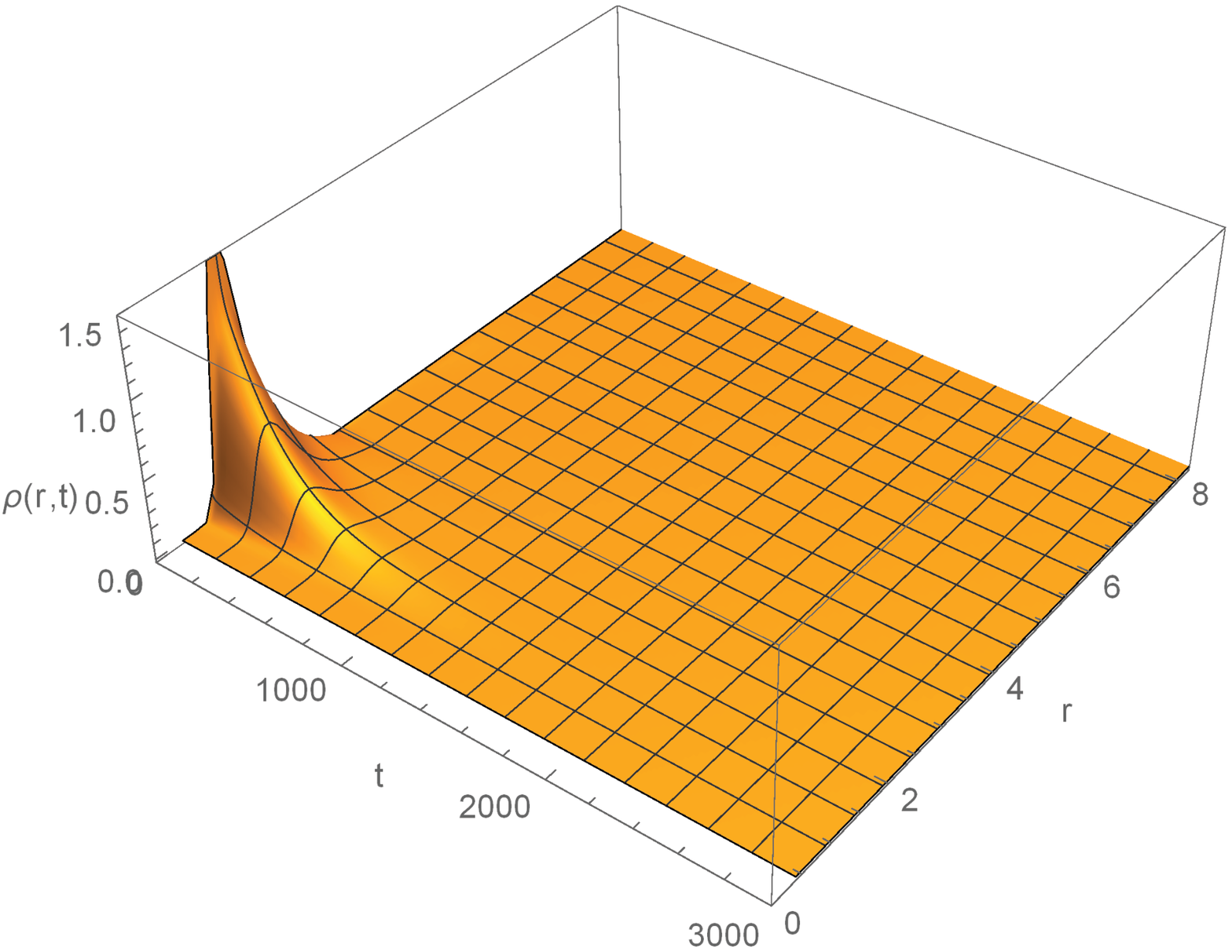}}}
 \caption{Time evolution of $\rho(r,t)$ in the first passage process of the RN-AdS black hole surrounded by quintessence with $a=0.1, T_E=0.02$. (a) $\omega_q=-2/3$ (b) $\omega_q=-2/3$ (c) $\omega_q=-0.9$ (d) $\omega_q=-0.9$. For the left two graphs, the initial condition is chosen as Gaussian wave pocket located at the large black hole state while for the right two graphs, the initial condition is chosen as Gaussian wave pocket located at the small black hole state.}
\label{fg3}
\end{figure}

In order to probe the dynamic evolution of $\rho(r,t)$ in the first passage process, we need to resolve the Fokker-Planck equation by adding the absorbing boundary condition ($\rho(r_m,t)=0$~\cite{liran2}) for the intermediate transition state(the Gibbs free energy achieves the local maximum here) besides the reflecting boundary condition mentioned in the former subsection. $r_m$ denotes the horizon radius corresponding to the intermediate transition state. Remind that we have solved the Fokker-Planck equation with only the reflecting boundary conditions in the former subsection. We plot the evolution of $\rho(r,t)$ in the first passage process for the RN-AdS black hole surrounded by quintessence with $a=0.1, T_E=0.02$ in Fig.\ref{fg3}. For the left two graphs, the initial condition is chosen as Gaussian wave pocket located at the large black hole state while for the right two graphs, the initial condition is chosen as Gaussian wave pocket located at the small black hole state. For comparison, the state parameter of quintessence dark energy $\omega_q$ is chosen as $-2/3$ and $-0.9$ respectively in the top and bottom rows. As shown in Fig.\ref{fg3}, the peaks located at the large (small) black hole decay very rapidly, irrespective of the initial black hole state. And the peaks of the case $\omega_q=-2/3$ decays more rapidly than that of the case $\omega_q=-0.9$, showing the effect of quintessence dark energy.

Suppose the probability that by time t the black hole system still stays at the large or small black hole state (not having accomplished a first passage) can be added up as $\Sigma(t)$, the distribution of first passage time (denoted as $F_p(t)$) reads~\cite{liran1}
\begin{equation}
F_p(t)=-\frac{d\Sigma(t)}{dt}.\label{13}
\end{equation}
Note that in the numerics we have considered $\Sigma(t)=\int^{r_m}_{0.3}\rho(r,t)dr$ for the initial state located at the small black hole while $\Sigma(t)=\int^{8.3}_{r_m}\rho(r,t)dr$ for the initial state located at the large black hole.

\begin{figure}[H]
\centerline{\subfigure[]{\label{w23a0.1T0.02L4}
\includegraphics[width=8cm,height=6cm]{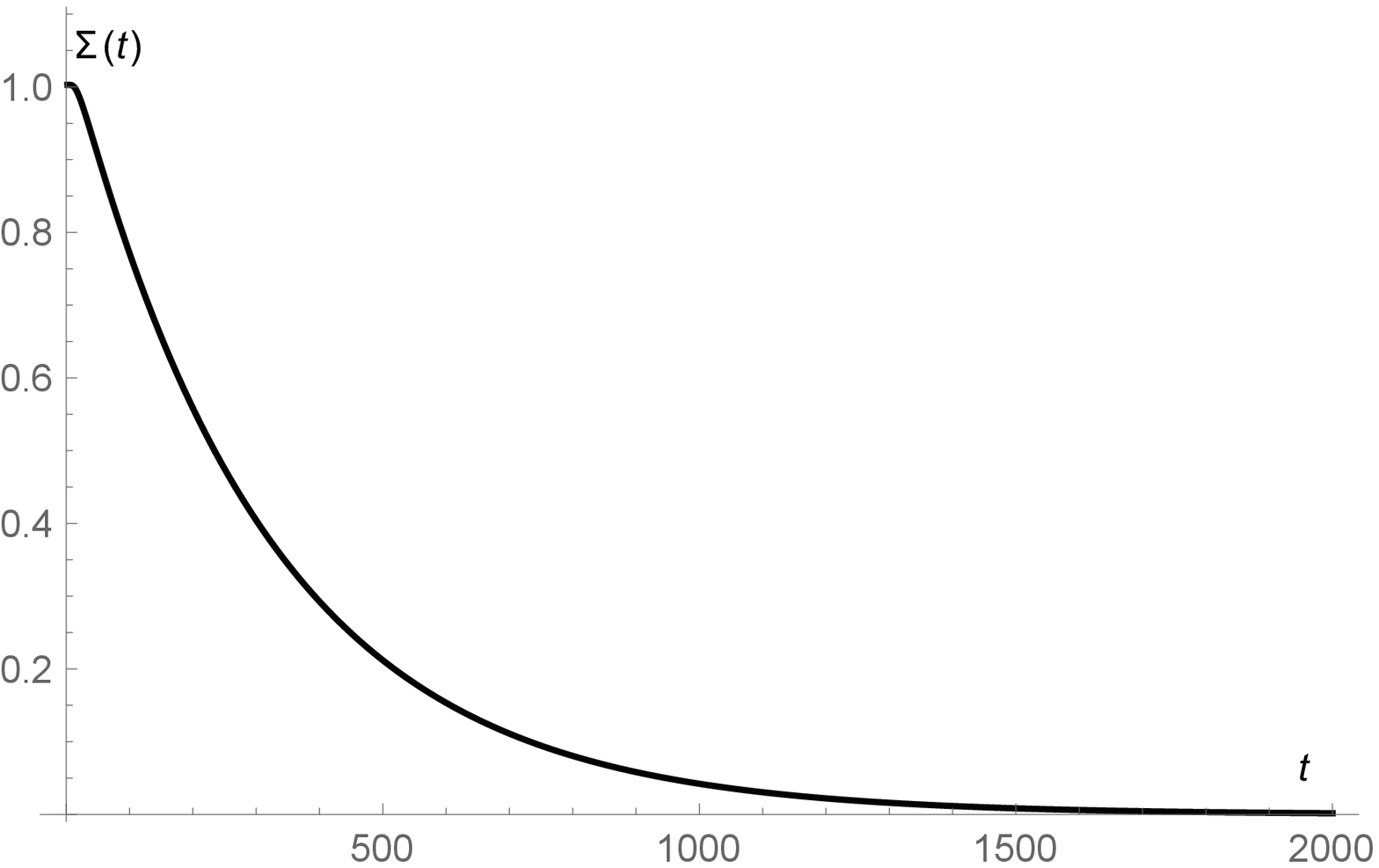}}
\subfigure[]{\label{w23a0.1T0.02S4}
\includegraphics[width=8cm,height=6cm]{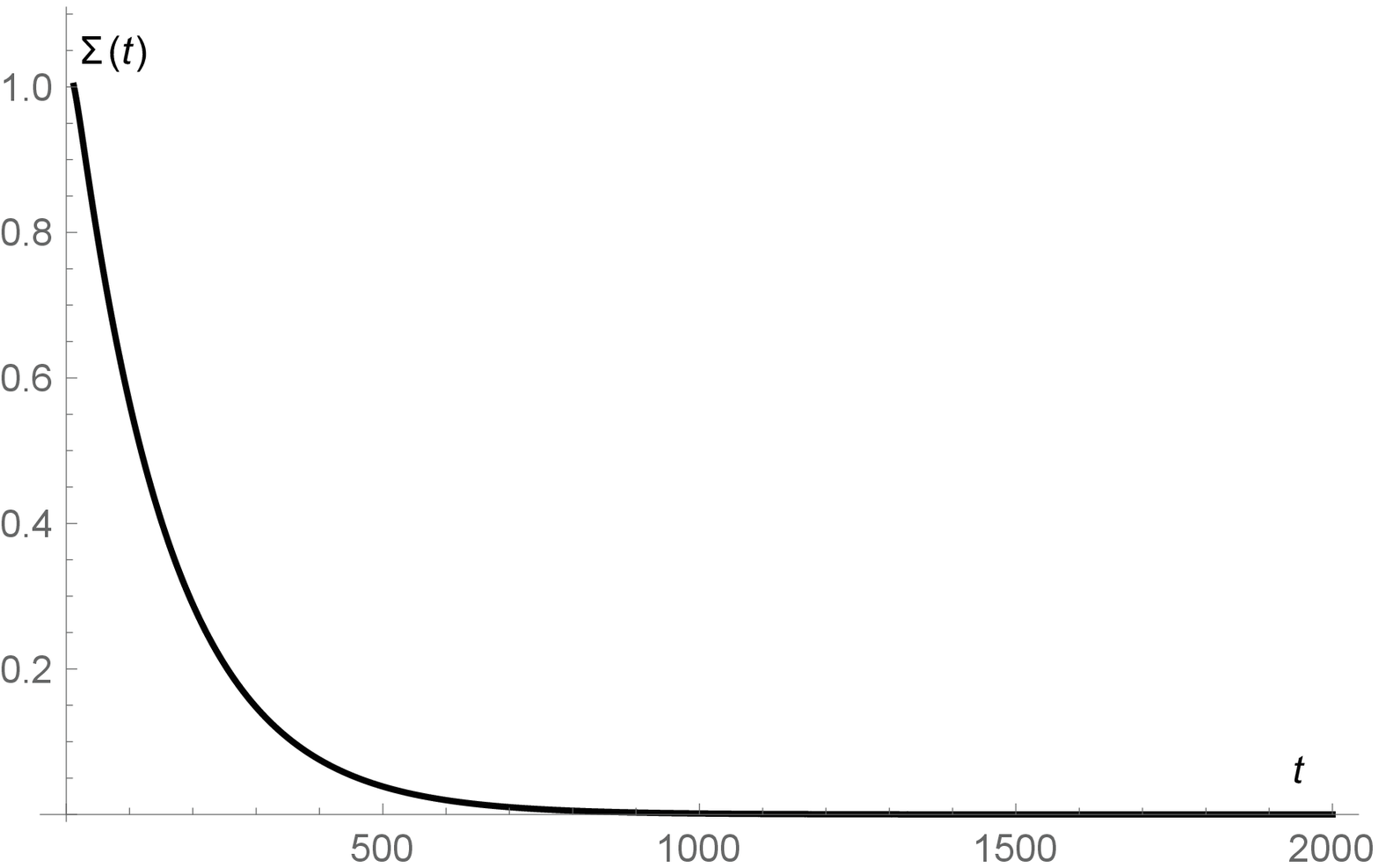}}}
\centerline{\subfigure[]{\label{w0.9a0.1T0.02L4}
\includegraphics[width=8cm,height=6cm]{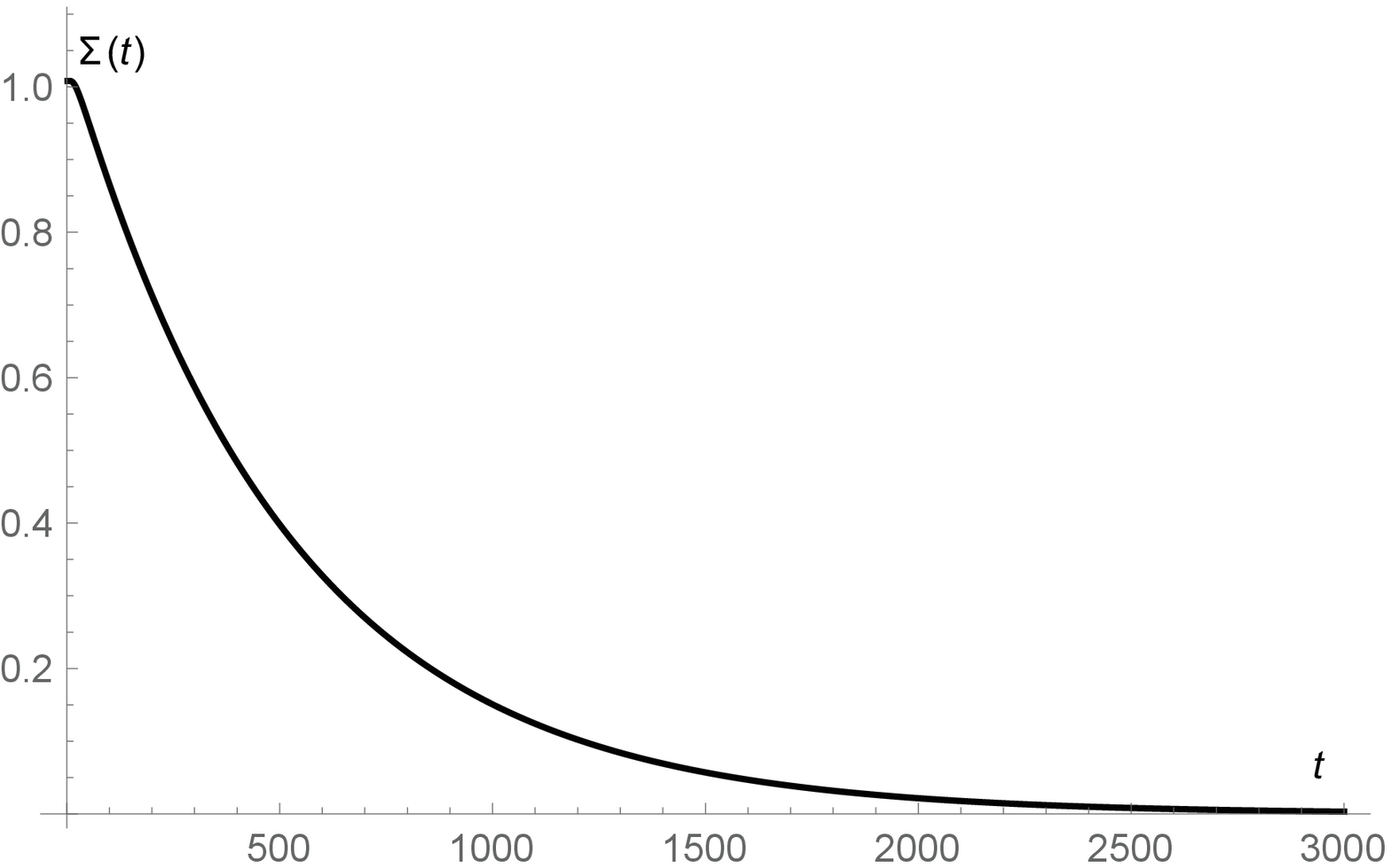}}
\subfigure[]{\label{w0.9a0.1T0.02S4}
\includegraphics[width=8cm,height=6cm]{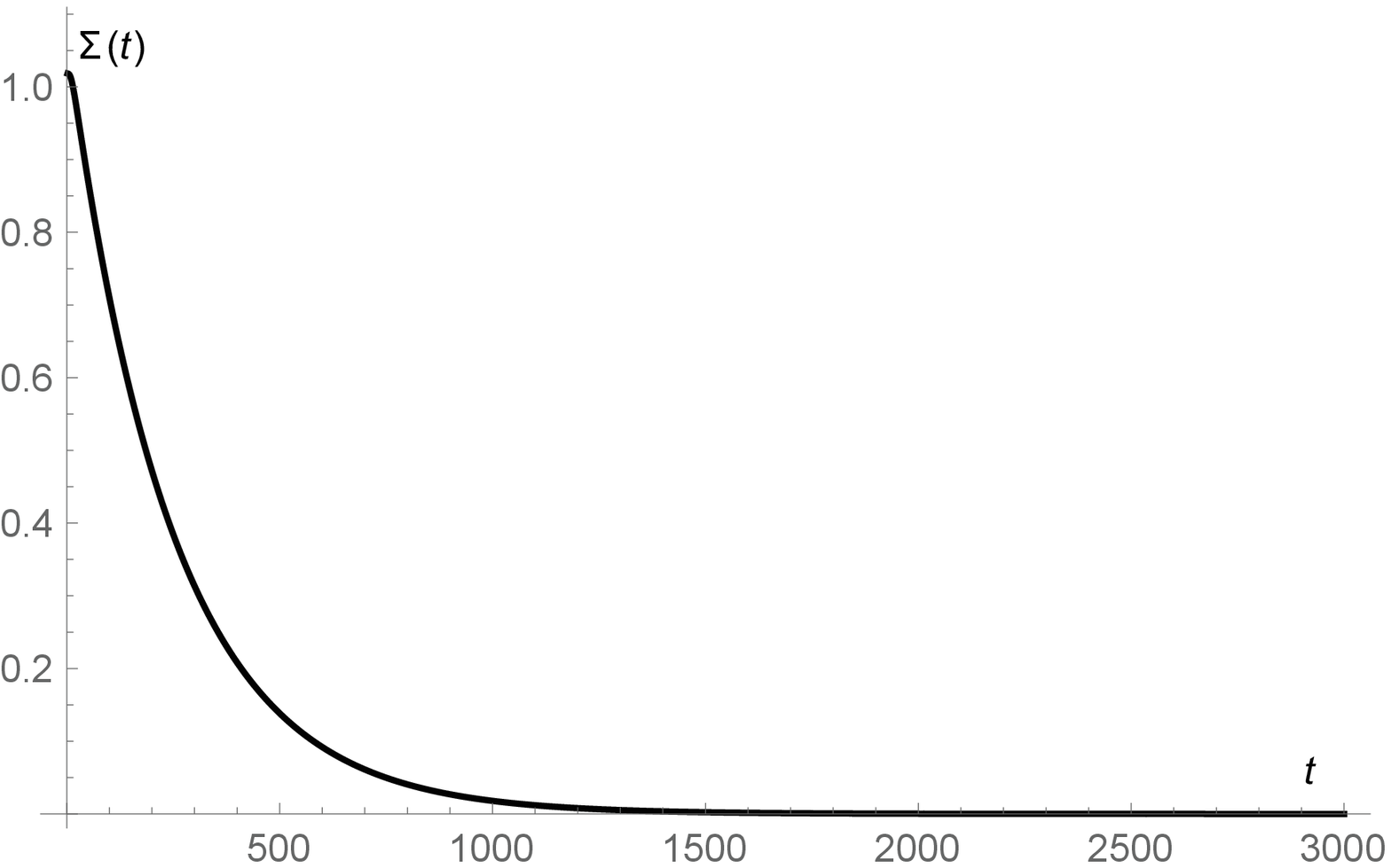}}}
 \caption{Time evolution of $\Sigma(t)$ for the RN-AdS black hole surrounded by quintessence with $a=0.1, T_E=0.02$. (a) $\omega_q=-2/3$ (b) $\omega_q=-2/3$ (c) $\omega_q=-0.9$ (d) $\omega_q=-0.9$. For the left two graphs, the initial condition is chosen as Gaussian wave pocket located at the large black hole state while for the right two graphs, the initial condition is chosen as Gaussian wave pocket located at the small black hole state.}
\label{fg4}
\end{figure}

The behavior of $\Sigma(t)$ is depicted in Fig.\ref{fg4}. No matter what the initial black hole state is, $\Sigma(t)$ decreases quickly, implying that the transition from the large (small) black hole state to the small (large) black hole state happens quickly. Consistent with the findings in Figs.\ref{fg1}-\ref{fg3}, the larger the state parameter of quintessence dark energy $\omega_q$ is, the faster $\Sigma(t)$ decreases.

\begin{figure}[H]
\centerline{\subfigure[]{\label{w23a0.1T0.02L5}
\includegraphics[width=8cm,height=6cm]{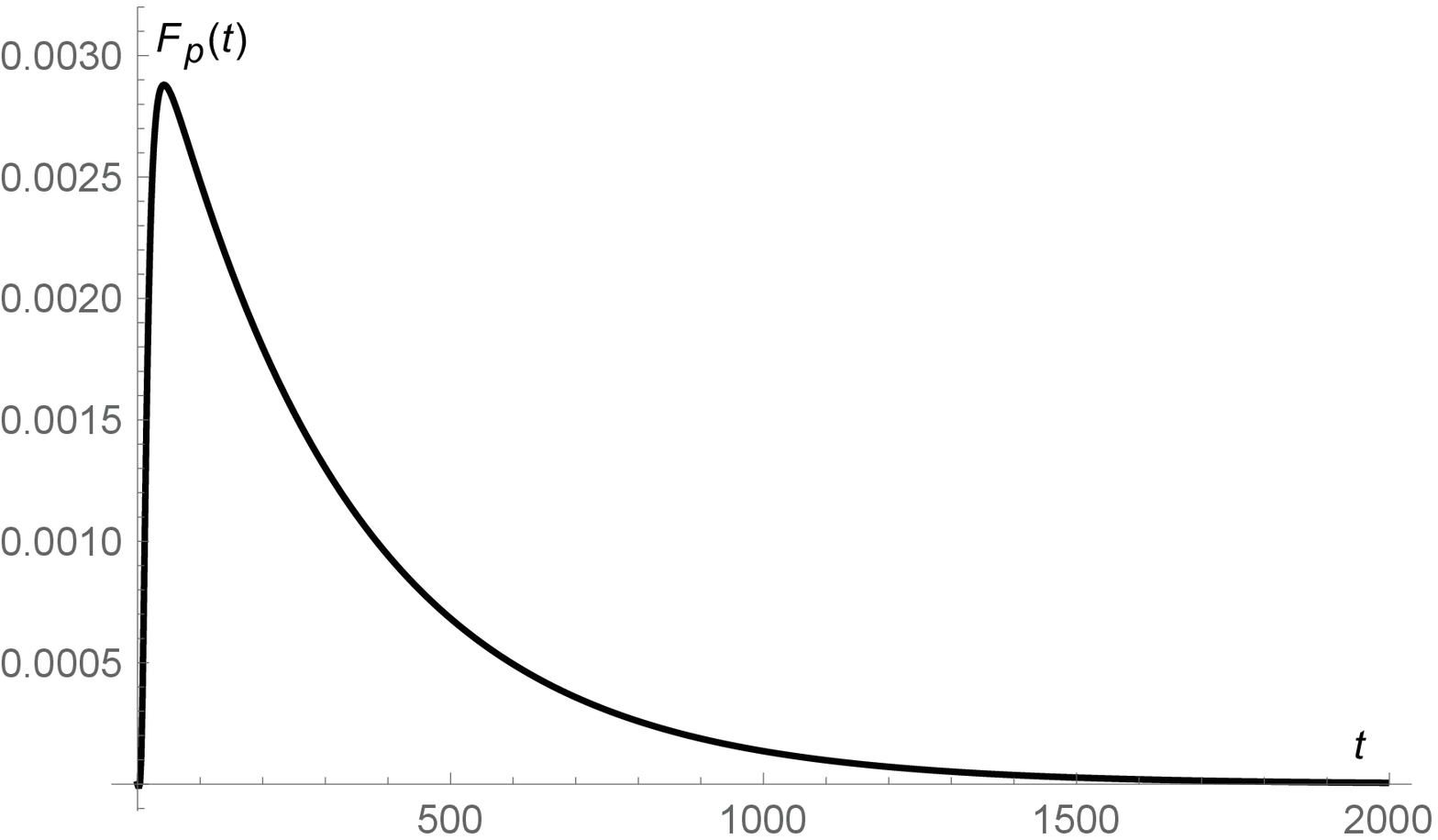}}
\subfigure[]{\label{w23a0.1T0.02S5}
\includegraphics[width=8cm,height=6cm]{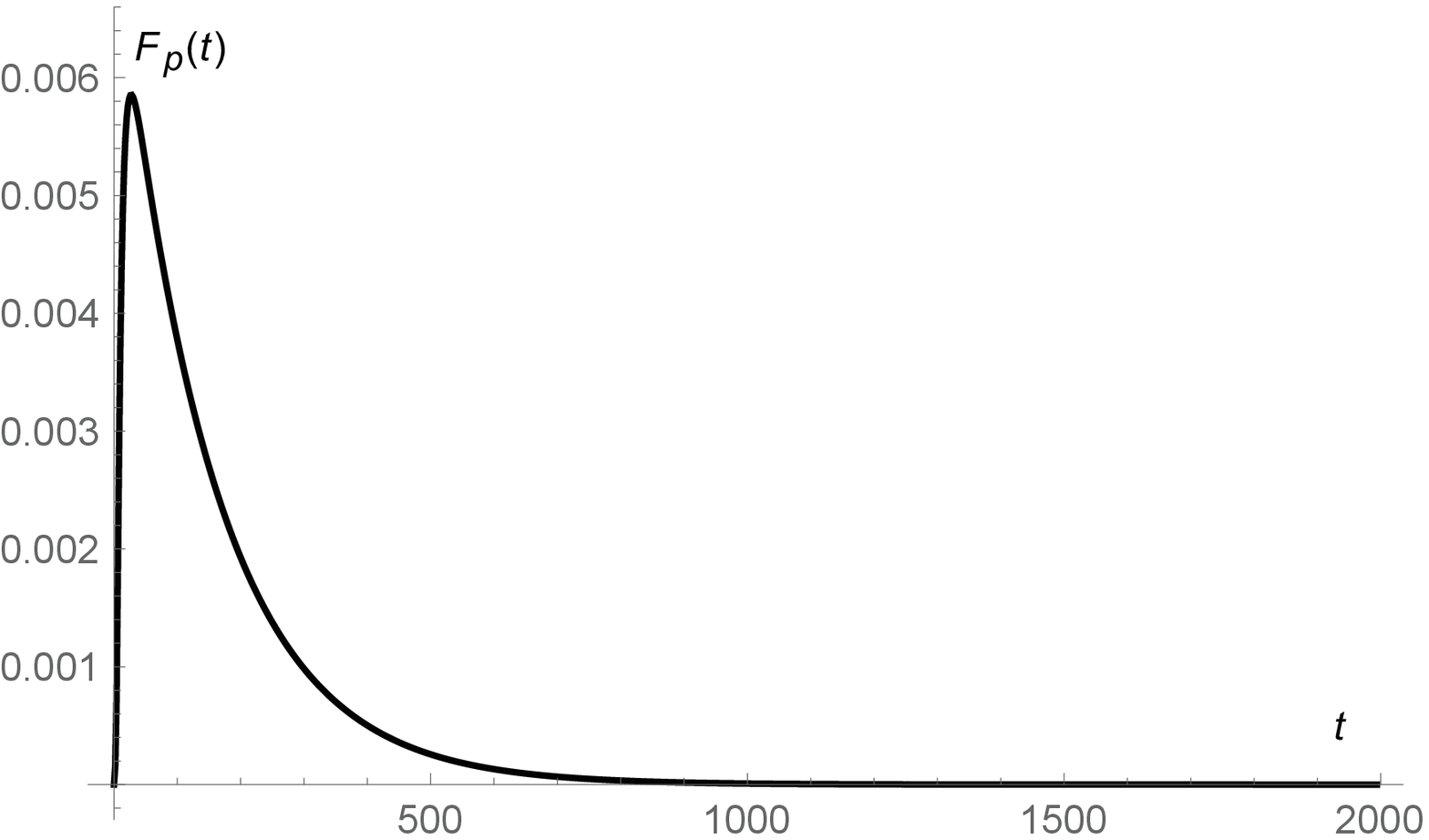}}}
\centerline{\subfigure[]{\label{w0.9a0.1T0.02L5}
\includegraphics[width=8cm,height=6cm]{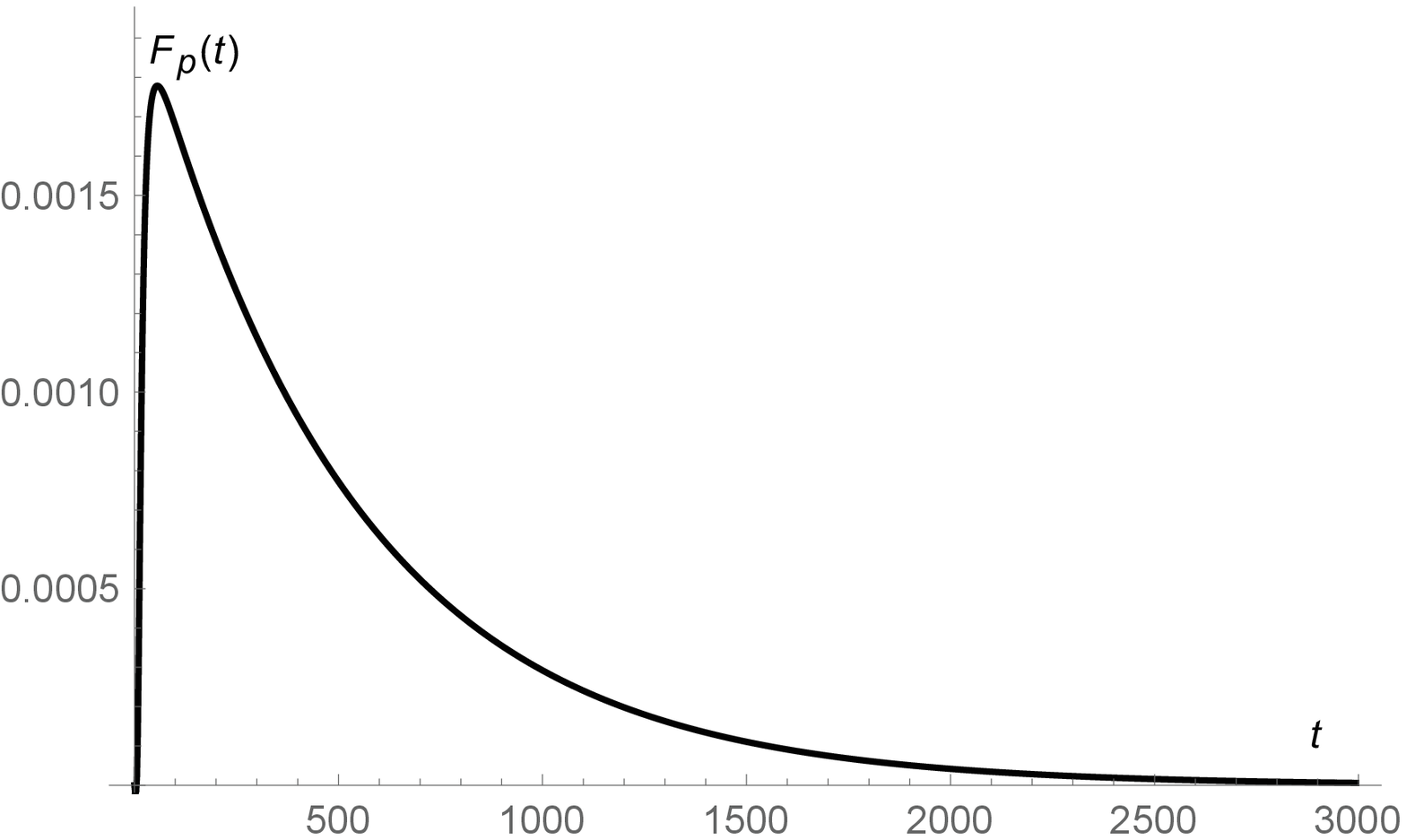}}
\subfigure[]{\label{w0.9a0.1T0.02S5}
\includegraphics[width=8cm,height=6cm]{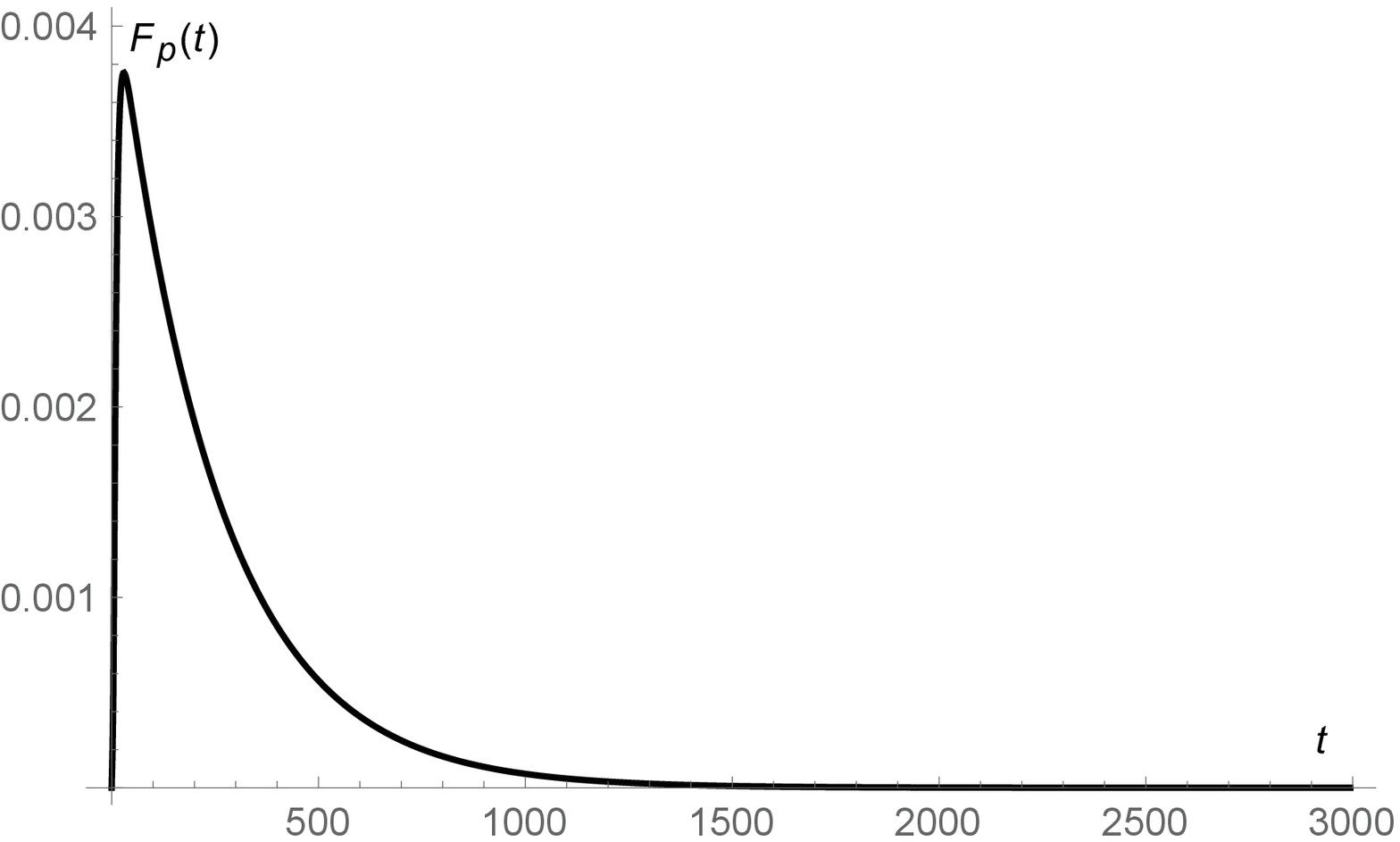}}}
 \caption{$F_p(t)$ for the RN-AdS black hole surrounded by quintessence with $a=0.1, T_E=0.02$. (a) $\omega_q=-2/3$ (b) $\omega_q=-2/3$ (c) $\omega_q=-0.9$ (d) $\omega_q=-0.9$. For the left two graphs, the initial condition is chosen as Gaussian wave pocket located at the large black hole state while for the right two graphs, the initial condition is chosen as Gaussian wave pocket located at the small black hole state.}
\label{fg5}
\end{figure}

Utilizing the Fokker-Planck equation and Eq. (\ref{13}), the relation between $F_p(t)$ and $\rho(r,t)$ has been derived as~\cite{liran2}
\begin{equation}
F_p(t)=-D\frac{\partial \rho(r,t)}{\partial r}\Big|_{r_m},\label{14}
\end{equation}
We plot $F_p(t)$ in Fig.\ref{fg5}. A single peak can be found in all the four cases, implying that considerable first passage events happen at short time.

\section{Conclusions}
\label{Sec4}
We investigate in detail the dynamic phase transition of charged AdS black holes surrounded by quintessence in this paper.

Firstly, we obtain the analytic expression of Gibbs free energy $G_L$ for charged AdS black holes surrounded by quintessence based on the Gibbs free energy landscape. We read off the phase transition temperature from the intersection point between the large black hole branch and the small black hole branch. Then we plot the $G_L-r_+$ curve at the phase transition temperature. As clearly shown in the $G_L-r_+$ curve, there exist double wells which have the same depth, providing further support on the former findings concerning the Gauss-Bonnet black holes~\cite{weishaowen2}.

Secondly, by numerically solving the Fokker-Planck equation with both the initial condition and reflecting boundary condition imposed, we probe the probabilistic evolution of charged AdS black holes surrounded by quintessence. We display the evolution of the probability distribution $\rho(r,t)$ graphically. We also plot the time evolution of both $\rho(r_l,t)$ and $\rho(r_s,t)$. These two sets of figures are mutually complementary with each other. For the case that the initial state is the large black hole, the peak denoting the large black hole gradually decreases while the other peak denoting the small black hole starts to grow from zero. And $\rho(r,t)$ approaches to be a stationary distribution in the long time limit with two peaks denoting the large and small black holes respectively. This clearly exhibits the dynamic process of how the initial large black hole evolves into the small black hole. More specifically, $\rho(r_l,t)$ is finite while $\rho(r_s,t)$ equals to zero at the begining. During the evolution, $\rho(r_l,t)$ decreases while $\rho(r_s,t)$ increases. Finally, they approach to the same value, consistent with the conclusion in Ref~\cite{weishaowen2} that the final probability of the large black hole equals to that of the small black hole because their Gibbs free energy equal to each other. Similar explanations is applicable when the initial black hole state is small black hole. Moreover, the effect of quintessence dark energy on the dynamic phase transition of black holes can be reflected in the above process. The initial black hole state evolves more quickly in the case of $\omega_q=-2/3$ than in the case $\omega_q=-0.9$.

Thirdly, we consider the first passage process which describes the process that the small (large) black hole state reaches the intermediate transition state for the first time. This first passage process can be interpreted intuitively from the Gibbs free energy landscape that for the first time the state at the well of $G_L$ reaches the state at the barriers of $G_L$. The corresponding time is called the first passage time. We resolve the Fokker-Planck equation by adding the absorbing boundary condition for the intermediate transition state. It is shown intuitively that the peaks located at the large (small) black hole decay very rapidly, irrespective of the initial black hole state. The larger the state parameter of quintessence dark energy $\omega_q$ is, the faster the corresponding peak decays, showing the effect of quintessence dark energy.

Fourthly, we discuss the quantity $\Sigma(t)$ which add up the probability that by time t the black hole system still stays at the large or small black hole state (not having accomplished a first passage). It is shown that $\Sigma(t)$ decreases quickly no matter what the initial black hole state is, suggesting that the transition from the large (small) black hole state to the small (large) black hole state happens quickly. It can also be witnessed from the distribution of the first passage time $F_p(t)$.

In one word, we have probed the kinetics of phase transition for charged AdS black holes surrounded by quintessence and disclose the effect of quintessence dark energy on the dynamic phase transition of charged AdS black hole.

 \section*{Acknowledgements}

 Shan-Quan Lan is supported by National Natural Science Foundation of China (Grant No.12005088). Gu-Qiang Li is supported by Guangdong Basic and Applied Basic Research Foundation, China (Grant Nos.2021A1515010246).

\end{document}